\documentclass[twocolumn,letterpaper,aps,prc,longbibliography,superscriptaddress,nofootinbib,floatfix]{revtex4-2}

\usepackage{graphicx}	
\usepackage{xspace}     
\usepackage{amsmath}     
\usepackage{xcolor}  
\begin{document}

\title{Measurements at forward rapidity of elliptic flow of charged 
hadrons and open-heavy-flavor muons in Au$+$Au collisions at 
$\sqrt{s_{_{NN}}}=200$ GeV}


\newcommand{\abilene}{Abilene Christian University, Abilene, Texas 79699, USA}
\newcommand{\augie}{Department of Physics, Augustana University, Sioux Falls, South Dakota 57197, USA}
\newcommand{\banaras}{Department of Physics, Banaras Hindu University, Varanasi 221005, India}
\newcommand{\barc}{Bhabha Atomic Research Centre, Bombay 400 085, India}
\newcommand{\baruch}{Baruch College, City University of New York, New York, New York, 10010 USA}
\newcommand{\bnlcoll}{Collider-Accelerator Department, Brookhaven National Laboratory, Upton, New York 11973-5000, USA}
\newcommand{\bnlphys}{Physics Department, Brookhaven National Laboratory, Upton, New York 11973-5000, USA}
\newcommand{\caucr}{University of California-Riverside, Riverside, California 92521, USA}
\newcommand{\charlesczech}{Charles University, Faculty of Mathematics and Physics, 180 00 Troja, Prague, Czech Republic}
\newcommand{\ciae}{Science and Technology on Nuclear Data Laboratory, China Institute of Atomic Energy, Beijing 102413, People's Republic of China}
\newcommand{\cns}{Center for Nuclear Study, Graduate School of Science, University of Tokyo, 7-3-1 Hongo, Bunkyo, Tokyo 113-0033, Japan}
\newcommand{\colorado}{University of Colorado, Boulder, Colorado 80309, USA}
\newcommand{\columbia}{Columbia University, New York, New York 10027 and Nevis Laboratories, Irvington, New York 10533, USA}
\newcommand{\czechtech}{Czech Technical University, Zikova 4, 166 36 Prague 6, Czech Republic}
\newcommand{\debrecen}{Debrecen University, H-4010 Debrecen, Egyetem t{\'e}r 1, Hungary}
\newcommand{\elte}{ELTE, E{\"o}tv{\"o}s Lor{\'a}nd University, H-1117 Budapest, P{\'a}zm{\'a}ny P.~s.~1/A, Hungary}
\newcommand{\ewha}{Ewha Womans University, Seoul 120-750, Korea}
\newcommand{\fsu}{Florida State University, Tallahassee, Florida 32306, USA}
\newcommand{\gsu}{Georgia State University, Atlanta, Georgia 30303, USA}
\newcommand{\hiroshima}{Physics Program and International Institute for Sustainability with Knotted Chiral Meta Matter (WPI-SKCM$^2$), Hiroshima University, Higashi-Hiroshima, Hiroshima 739-8526, Japan}
\newcommand{\howard}{Department of Physics and Astronomy, Howard University, Washington, DC 20059, USA}
\newcommand{\hunrenatomki}{HUN-REN ATOMKI, H-4026 Debrecen, Bem t{\'e}r 18/c, Hungary}
\newcommand{\ihepprot}{IHEP Protvino, State Research Center of Russian Federation, Institute for High Energy Physics, Protvino, 142281, Russia}
\newcommand{\illuiuc}{University of Illinois at Urbana-Champaign, Urbana, Illinois 61801, USA}
\newcommand{\inrras}{Institute for Nuclear Research of the Russian Academy of Sciences, prospekt 60-letiya Oktyabrya 7a, Moscow 117312, Russia}
\newcommand{\instpasczech}{Institute of Physics, Academy of Sciences of the Czech Republic, Na Slovance 2, 182 21 Prague 8, Czech Republic}
\newcommand{\isu}{Iowa State University, Ames, Iowa 50011, USA}
\newcommand{\jaea}{Advanced Science Research Center, Japan Atomic Energy Agency, 2-4 Shirakata Shirane, Tokai-mura, Naka-gun, Ibaraki-ken 319-1195, Japan}
\newcommand{\jeonbuk}{Jeonbuk National University, Jeonju, 54896, Korea}
\newcommand{\jyvaskyla}{Helsinki Institute of Physics and University of Jyv{\"a}skyl{\"a}, P.O.Box 35, FI-40014 Jyv{\"a}skyl{\"a}, Finland}
\newcommand{\kek}{KEK, High Energy Accelerator Research Organization, Tsukuba, Ibaraki 305-0801, Japan}
\newcommand{\korea}{Korea University, Seoul 02841, Korea}
\newcommand{\kurchatov}{National Research Center ``Kurchatov Institute", Moscow, 123098 Russia}
\newcommand{\kyoto}{Kyoto University, Kyoto 606-8502, Japan}
\newcommand{\lawllnl}{Lawrence Livermore National Laboratory, Livermore, California 94550, USA}
\newcommand{\losalamos}{Los Alamos National Laboratory, Los Alamos, New Mexico 87545, USA}
\newcommand{\lund}{Department of Physics, Lund University, Box 118, SE-221 00 Lund, Sweden}
\newcommand{\lyon}{IPNL, CNRS/IN2P3, Univ Lyon, Universit{\'e} Lyon 1, F-69622, Villeurbanne, France}
\newcommand{\maryland}{University of Maryland, College Park, Maryland 20742, USA}
\newcommand{\mass}{Department of Physics, University of Massachusetts, Amherst, Massachusetts 01003-9337, USA}
\newcommand{\mate}{MATE, Laboratory of Femtoscopy, K\'aroly R\'obert Campus, H-3200 Gy\"ongy\"os, M\'atrai \'ut 36, Hungary}
\newcommand{\michigan}{Department of Physics, University of Michigan, Ann Arbor, Michigan 48109-1040, USA}
\newcommand{\miss}{Mississippi State University, Mississippi State, Mississippi 39762, USA}
\newcommand{\muhlenberg}{Muhlenberg College, Allentown, Pennsylvania 18104-5586, USA}
\newcommand{\nara}{Nara Women's University, Kita-uoya Nishi-machi Nara 630-8506, Japan}
\newcommand{\natmephi}{National Research Nuclear University, MEPhI, Moscow Engineering Physics Institute, Moscow, 115409, Russia}
\newcommand{\newmex}{University of New Mexico, Albuquerque, New Mexico 87131, USA}
\newcommand{\nmsu}{New Mexico State University, Las Cruces, New Mexico 88003, USA}
\newcommand{\northcg}{Physics and Astronomy Department, University of North Carolina at Greensboro, Greensboro, North Carolina 27412, USA}
\newcommand{\ohio}{Department of Physics and Astronomy, Ohio University, Athens, Ohio 45701, USA}
\newcommand{\ornl}{Oak Ridge National Laboratory, Oak Ridge, Tennessee 37831, USA}
\newcommand{\orsay}{IPN-Orsay, Univ.~Paris-Sud, CNRS/IN2P3, Universit\'e Paris-Saclay, BP1, F-91406, Orsay, France}
\newcommand{\peking}{Peking University, Beijing 100871, People's Republic of China}
\newcommand{\pnpi}{PNPI, Petersburg Nuclear Physics Institute, Gatchina, Leningrad region, 188300, Russia}
\newcommand{\pusan}{Pusan National University, Pusan 46241, Korea}
\newcommand{\riken}{RIKEN Nishina Center for Accelerator-Based Science, Wako, Saitama 351-0198, Japan}
\newcommand{\rikjrbrc}{RIKEN BNL Research Center, Brookhaven National Laboratory, Upton, New York 11973-5000, USA}
\newcommand{\rikkyo}{Physics Department, Rikkyo University, 3-34-1 Nishi-Ikebukuro, Toshima, Tokyo 171-8501, Japan}
\newcommand{\saispbstu}{Saint Petersburg State Polytechnic University, St.~Petersburg, 195251 Russia}
\newcommand{\seoulnat}{Department of Physics and Astronomy, Seoul National University, Seoul 151-742, Korea}
\newcommand{\stonybrkc}{Chemistry Department, Stony Brook University, SUNY, Stony Brook, New York 11794-3400, USA}
\newcommand{\stonycrkp}{Department of Physics and Astronomy, Stony Brook University, SUNY, Stony Brook, New York 11794-3800, USA}
\newcommand{\tenn}{University of Tennessee, Knoxville, Tennessee 37996, USA}
\newcommand{\titech}{Department of Physics, Tokyo Institute of Technology, Oh-okayama, Meguro, Tokyo 152-8551, Japan}
\newcommand{\tsukuba}{Tomonaga Center for the History of the Universe, University of Tsukuba, Tsukuba, Ibaraki 305, Japan}
\newcommand{\usmma}{United States Merchant Marine Academy, Kings Point, New York 11024, USA}
\newcommand{\vandy}{Vanderbilt University, Nashville, Tennessee 37235, USA}
\newcommand{\weizmann}{Weizmann Institute, Rehovot 76100, Israel}
\newcommand{\wigner}{Institute for Particle and Nuclear Physics, HUN-REN Wigner Research Centre for Physics, (HUN-REN Wigner RCP, RMI), H-1525 Budapest 114, POBox 49, Budapest, Hungary}
\newcommand{\yonsei}{Yonsei University, IPAP, Seoul 120-749, Korea}
\newcommand{\zagreb}{Department of Physics, Faculty of Science, University of Zagreb, Bijeni\v{c}ka c.~32 HR-10002 Zagreb, Croatia}
\newcommand{\zambia}{Department of Physics, School of Natural Sciences, University of Zambia, Great East Road Campus, Box 32379, Lusaka, Zambia}
\affiliation{\abilene}
\affiliation{\augie}
\affiliation{\banaras}
\affiliation{\barc}
\affiliation{\baruch}
\affiliation{\bnlcoll}
\affiliation{\bnlphys}
\affiliation{\caucr}
\affiliation{\charlesczech}
\affiliation{\ciae}
\affiliation{\cns}
\affiliation{\colorado}
\affiliation{\columbia}
\affiliation{\czechtech}
\affiliation{\debrecen}
\affiliation{\elte}
\affiliation{\ewha}
\affiliation{\fsu}
\affiliation{\gsu}
\affiliation{\hiroshima}
\affiliation{\howard}
\affiliation{\hunrenatomki}
\affiliation{\ihepprot}
\affiliation{\illuiuc}
\affiliation{\inrras}
\affiliation{\instpasczech}
\affiliation{\isu}
\affiliation{\jaea}
\affiliation{\jeonbuk}
\affiliation{\jyvaskyla}
\affiliation{\kek}
\affiliation{\korea}
\affiliation{\kurchatov}
\affiliation{\kyoto}
\affiliation{\lawllnl}
\affiliation{\losalamos}
\affiliation{\lund}
\affiliation{\lyon}
\affiliation{\maryland}
\affiliation{\mass}
\affiliation{\mate}
\affiliation{\michigan}
\affiliation{\miss}
\affiliation{\muhlenberg}
\affiliation{\nara}
\affiliation{\natmephi}
\affiliation{\newmex}
\affiliation{\nmsu}
\affiliation{\northcg}
\affiliation{\ohio}
\affiliation{\ornl}
\affiliation{\orsay}
\affiliation{\peking}
\affiliation{\pusan} 
\affiliation{\pnpi}
\affiliation{\riken}
\affiliation{\rikjrbrc}
\affiliation{\rikkyo}
\affiliation{\saispbstu}
\affiliation{\seoulnat}
\affiliation{\stonybrkc}
\affiliation{\stonycrkp}
\affiliation{\tenn}
\affiliation{\titech}
\affiliation{\tsukuba}
\affiliation{\usmma}
\affiliation{\vandy}
\affiliation{\weizmann}
\affiliation{\wigner}
\affiliation{\yonsei}
\affiliation{\zagreb}
\affiliation{\zambia}
\author{N.J.~Abdulameer} \affiliation{\debrecen} \affiliation{\hunrenatomki}
\author{U.~Acharya} \affiliation{\gsu}
\author{A.~Adare} \affiliation{\colorado} 
\author{C.~Aidala} \affiliation{\michigan} 
\author{N.N.~Ajitanand} \altaffiliation{Deceased} \affiliation{\stonybrkc} 
\author{Y.~Akiba} \email[PHENIX Spokesperson: ]{akiba@rcf.rhic.bnl.gov} \affiliation{\riken} \affiliation{\rikjrbrc}
\author{M.~Alfred} \affiliation{\howard} 
\author{S.~Antsupov} \affiliation{\saispbstu}
\author{K.~Aoki} \affiliation{\kek} \affiliation{\riken} 
\author{N.~Apadula} \affiliation{\isu} \affiliation{\stonycrkp} 
% ...
\collaboration{PHENIX Collaboration}  \noaffiliation

\date{\today}


\begin{abstract}


We present the first forward-rapidity measurements of elliptic 
anisotropy of open-heavy-flavor muons at the Relativistic Heavy Ion 
Collider. The measurements are based on data samples of Au$+$Au 
collisions at $\sqrt{s_{_{NN}}}=200$ GeV collected by the PHENIX 
experiment in 2014 and 2016 with integrated luminosity of 
$14.5~{\rm nb}^{-1}$. The measurements are performed in the pseudorapidity 
range $1.2<|\eta|<2$ and cover transverse momenta $1<p_T<4$~GeV/$c$. 
The elliptic flow of charged hadrons as a function of transverse 
momentum is also measured in the same kinematic range. We observe 
significant elliptic flow for both charged hadrons and heavy-flavor 
muons. The results show clear mass ordering of elliptic flow of light- 
and heavy-flavor particles. The magnitude of the measured $v_2$ is 
comparable to that in the midrapidity region. This indicates that there 
is no strong longitudinal dependence in the quark-gluon-plasma 
evolution between midrapidity and the rapidity range of this 
measurement at $\sqrt{s_{_{NN}}}=200$~GeV.

\end{abstract}

\maketitle

\section{Introduction}
\label{sec:Intro}

The quark-gluon plasma (QGP) formed in high-energy heavy ion 
collisions~\cite{PHENIX:2004vcz,BRAHMS:2004adc,PHOBOS:2004zne,STAR:2005gfr} 
is a hot, dense state of matter comprised of strongly 
interacting quarks and gluons. A key property of QGP is its 
near-perfect-fluid behavior characterized by the azimuthal anisotropy 
of the produced particles, which is correlated with anisotropies in the 
initial-state energy density~\cite{Shuryak:2004cy}. The elliptic 
azimuthal anisotropy ($v_2$) is defined by the amplitude of the second 
harmonic in a Fourier series expansion of the azimuthal distribution of 
produced particles,

\begin{equation}
\frac{dN}{d\phi} \propto 1+\sum_{n} 2v_n \cos[n(\phi - \Psi_n)]
\end{equation}

\noindent where $v_n = \langle \cos[n(\phi - \Psi_n)]\rangle$ is the 
n$^{th}$ order flow coefficient, $\phi$ is the azimuthal angle of any 
given particle, and $\Psi_n$ is the angle of the n$^{th}$-order 
symmetry plane.

Theoretical approaches involve quantum chomodynamics (QCD). Of 
particular interest is the azimuthal anisotropy of particles containing 
charm and beauty quarks. Because these heavy quarks have masses greater 
than the initial temperature of the QGP and the $\Lambda_{QCD}$ scale, 
they are produced in the initial parton-parton hard scatterings and 
their production can be described~\cite{Cacciari:2005rk} by 
perturbative QCD.  Once produced, the heavy quarks cannot be 
destroyed by the QGP and will experience the whole evolution of the 
system, giving access to QGP transport 
properties~\cite{annurev:annurev-nucl-101918-023806}.

Previous PHENIX measurements of cross sections of electrons from 
separated charm and beauty decays suggest a stronger suppression of 
charmed hadrons than for beauty 
hadrons~\cite{PHENIX:2015ynp,PHENIX:2022wim}. This indicates a mass 
ordering of the quark interactions with the QGP medium. The specific 
mechanisms that contribute to how heavy quarks interact with the QGP 
are still not entirely understood. Different effects such as 
radiative~\cite{Mustafa:1997pm,Dokshitzer:2001zm} and 
collisional~\cite{Meistrenko:2012ju} energy loss dominate the 
interactions depending on the heavy-quark transverse momentum. 
Low-p$_T$ charm and beauty quarks relevant to our measurement are 
nonrelativistic, and their propagation and interaction with the medium 
gives access to the drag and diffusion-transport coefficients of the 
QGP~\cite{Cao:2018ews}. It is important to understand if and how heavy 
quarks equilibrate with the QGP, and how hadronization via 
coalescence~\cite{He:2019vgs} of light and heavy quarks may influence 
the flow of the final-state hadrons. Measurements of collective flow of 
light and heavy hadrons in different kinematic regions may help answer 
these questions.

Much work has been done at the Relativistic Heavy Ion Collider (RHIC) 
to study the azimuthal anisotropy of heavy-flavor particles at 
midrapidity~\cite{PHENIX:2006iih,PHENIX:2015ynp,STAR:2017kkh}. These 
previous RHIC results show a nonzero elliptic flow for heavy-flavor 
particles and a clear mass ordering between heavy and light hadrons. 
However, there have been no RHIC measurements of elliptic flow of 
heavy-flavor particles at forward rapidity. Particles produced at 
forward rapidity sample a different area of phase space than those 
produced at midrapidity and could also be subject to different pressure 
gradients and temperatures within the QGP.  This makes forward-rapidity 
measurements an important part of understanding particle interactions 
with the medium. At the Large Hadron Collider (LHC), the forward 
rapidity ($1<|\eta|<2$) and midrapidity measurements($|\eta|<1$) of 
open-heavy-flavor $v_2$~\cite{CMS:2020bnz} show a similar trend. 
However, the boost-invariant region of QGP is much wider at the LHC 
than at RHIC due to the higher center-of-mass energies. Heavy-flavor 
production rates and the QGP temperature are significantly different 
between RHIC and the LHC. Therefore, these forward-rapidity 
measurements at RHIC energies are needed.

The organization of this paper is as follows. Section~\ref{sec:PHENIX} 
describes the PHENIX detector with special detail paid to the 
forward-rapidity detectors that are crucial for this analysis. 
Sections~\ref{sec:Methods} and~\ref{sec:Procedure} describe the 
experimental methods and procedure. Section~\ref{sec:Results} presents 
the first measurements at RHIC energies of the $v_2$ of decay muons 
from open-heavy-flavor mesons at forward rapidity and the measurement 
of the $v_2$ of charged hadrons. Finally, Section~\ref{sec:Conclusion} 
presents conclusions and commentary on the results.

\section{PHENIX detector}
\label{sec:PHENIX}

The PHENIX detector~\cite{PHENIX:1998vmi} comprises the global 
detectors, the central-arm spectrometers, and the muon-arm 
spectrometers. The global detectors (the zero-degree calorimeters and 
the beam-beam counters (BBCs)~\cite{ALLEN2003549}) are used to define a 
minimum-bias trigger as well as to determine the primary vertex and 
charged-particle multiplicity~\cite{ALLEN2003549}.

\begin{figure}[htb]
\includegraphics[width=1.0\linewidth]{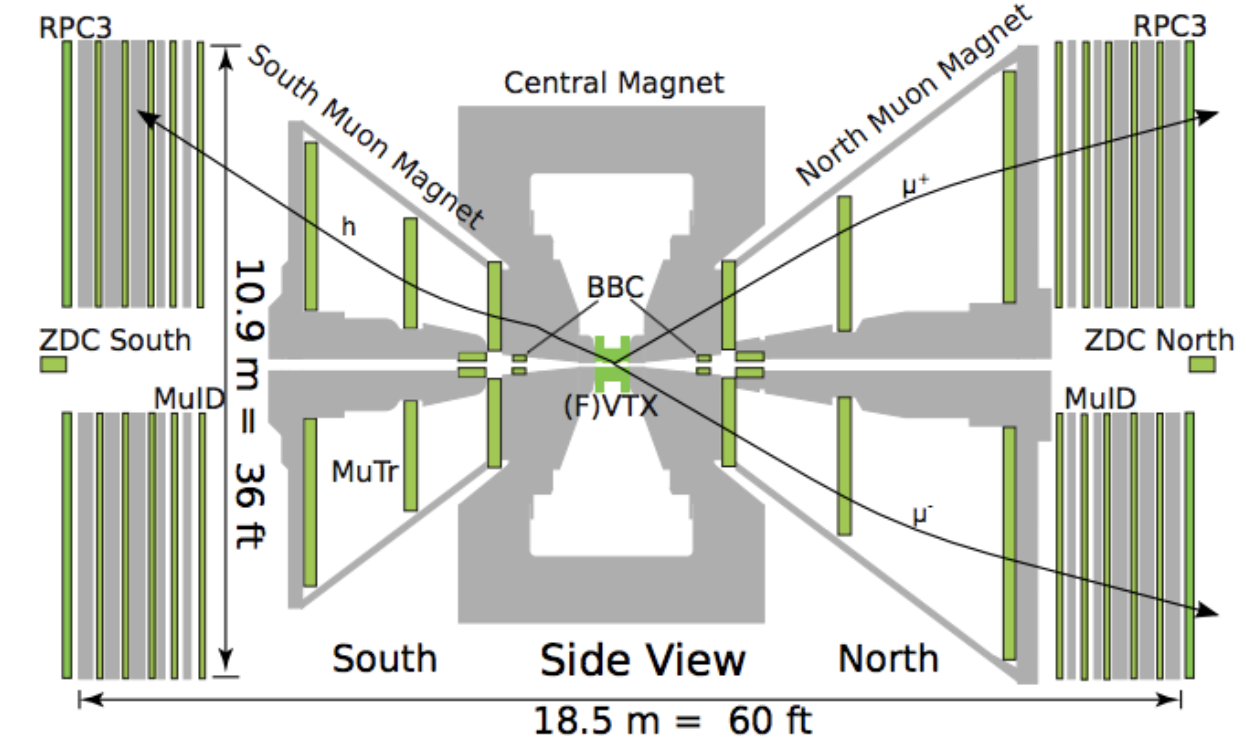}
\caption{Side view of the PHENIX detector.}
\label{fig:PHENIX}
\end{figure}

The central-arm spectrometers provide precise tracking and particle 
identification for electrons, charged hadrons, and photons. Tracking 
and momentum determination are provided by drift chambers and pad 
chambers. Particle identification is provided by time of flight, 
ring-imaging \v{C}erenkov detectors, and electromagnetic calorimeters. 
The BBCs and the central-arm tracking detectors are used for 
event-plane (EP) determination along with the forward-silicon-vertex 
detectors (FVTX), as described in Section~\ref{sec:Methods}. The 
central-arm silicon-vertex detector (VTX)~\cite{Nouicer:2007rb} is used 
to precisely determine the primary vertex point along the beam axis. 
Figure~\ref{fig:PHENIX} shows a side view of the PHENIX detector with 
the muon-arm spectrometers, BBC, (F)VTX, and the central magnet.

\subsection{The Muon-arm Spectrometers}
\label{subsec:Muonarms}

The PHENIX muon-arm spectrometers cover a pseudorapidity range of 
$1.2<|\eta|<2.2$ and full $2\pi$ in azimuth. Each muon spectrometer 
comprises hadron-absorbing material, muon trackers (MuTr) inside a 
conical-shaped magnet, and a muon identifier 
(MuID)~\cite{Akikawa:2003zs}. Separate from the spectrometers, but 
crucial for this analysis, is the forward-silicon-vertex detector 
(FVTX)~\cite{AIDALA201444}. The first layer of hadron-absorber material 
is placed between the FVTX and MuTr comprising 19 cm of copper, 
followed by 60 cm of iron from the central magnet, and then 36.2 cm of 
steel, all of which correspond to 7.2 nuclear-interaction lengths. This 
absorbing material is meant to block pions and kaons emitted into the 
acceptance range of the spectrometers, but allow muons to penetrate. 
Tracking and momentum measurements are performed by the MuTr that 
comprises eight octants of cathode-strip chambers in three different 
stations. The momentum resolution achieved by the MuTr is approximately 
$\delta p/p=0.05$ for muons. The MuID system is located downstream of 
the MuTr and contains five absorber plates in addition to the absorber 
material before the MuTr. Two Iarocci tube planes with vertical and 
horizontal orientations are placed after each of the five absorber 
plates. The general operating principle is that hadrons that manage to 
penetrate the initial absorbing layers between the FVTX and MuTr will 
not penetrate all of the MuID absorption plates (hadron-rich in Fig. 
2), whereas muons will penetrate all the plates and leave hits in the 
final set of Iarocci tubes (muon-rich in Fig. 2).

\subsection{Silicon Vertex Detectors (F)VTX}

The silicon-vertex detector (VTX) has four radial layers placed at 2.6, 
5.1, 11.8, 16.7 cm from the z-axis covering 
2$\times\Delta\phi{\approx}1.6\pi$ and $|z_{\rm vtx}|<10$ cm. The 
forward VTX detector (FVTX)~\cite{AIDALA201444} is a silicon-strip 
detector comprising four layers at 20.1, 26.1, 32.2, and 38.2 cm along 
the z-axis that measure particle trajectories with an $\eta$ acceptance 
of $1.2<|\eta|<2.2$. The FVTX comprises 96 radial strips covering 
full $2\pi$ in azimuth, providing a single-hit resolution of 25 
$\mu$m. The VTX and FVTX measure the radial distance of closest 
approach (${\rm DCA}_R$, which is determined by projecting the particle 
track from the FVTX onto a plane perpendicular to the z-axis located at 
the primary vertex point determined by the VTX.  Precisely measuring 
${\rm DCA}_R$ allows for a statistical separation of muons from 
short-lived particles, like heavy-flavor mesons, and those from 
long-lived particles such as pions and kaons. The {\sc pythia}+{\sc 
geant4} simulations show that selecting particles with 
$|{\rm DCA}_R|<500$ $\mu$m (Fig.~\ref{fig:DCAR}) greatly increases the 
purity of the measured sample of heavy-flavor muons, as well as 
rejecting secondary hadrons.

\begin{figure*} [ht]
\includegraphics[width=0.99\linewidth]{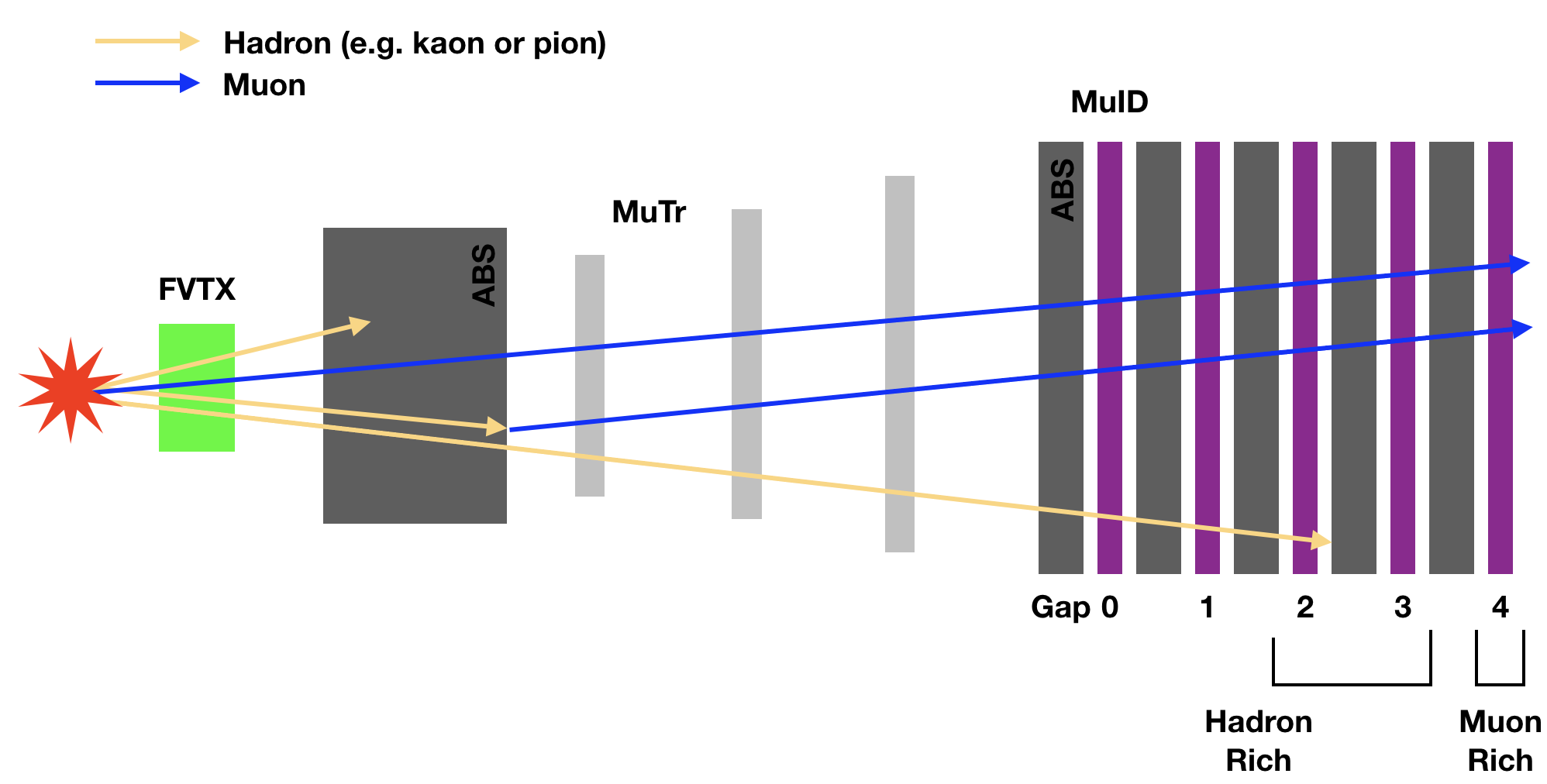}
\caption{Schematic of hadron and muon propagation through the PHENIX 
FVTX and muon spectrometers. Muons will penetrate all absorber layers 
leaving hits in the FVTX, MuTr, and final layer of the MuID, whereas 
hadrons will be stopped by one of the absorber layers.}
\label{fig:muonSpec}
\end{figure*}

\begin{figure}
\includegraphics[width=1.0\linewidth]{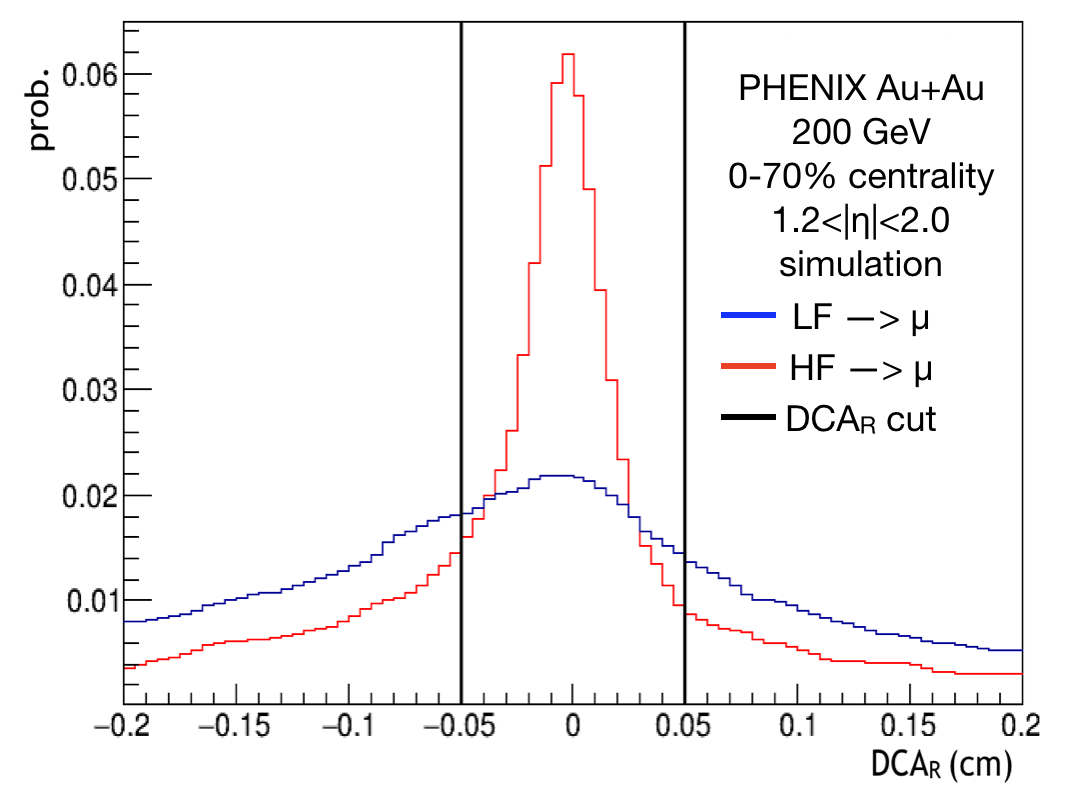}
\caption{{\sc pythia}+{\sc geant4} simulation of decay muons from light 
(LF) and heavy-flavor (HF) hadrons. Both distributions are normalized 
to one and we see that at ${\rm DCA}_R$ values close to zero 
($|{\rm DCA}_R|<0.05$ cm) muons from heavy-flavor hadrons are enriched 
relative to those from light hadrons.}
\label{fig:DCAR}
\end{figure}

\section{Analysis Methods}
\label{sec:Methods}

The PHENIX collaboration studied Au$+$Au collisions at 200 GeV using 
data collected at RHIC in 2014 and 2016 with a combined integrated 
luminosity of $14.5~{\rm nb}^{-1}$. The analysis was performed for 
charged hadrons and muons over the transverse momentum range $1<p_T<4$ 
GeV/$c$ with tracks being required to have multiple hits in the FVTX, 
MuTr, and MuID as described below. A large fraction of the tracks in 
the muon spectrometer come from secondary particles produced in the 
hadronic absorber. This contamination is largely suppressed by 
requiring that tracks in the MuTr match with tracks in the FVTX. 
However, it is common for a single MuTr track to be associated with 
multiple FVTX tracks due to the large cone projection caused by 
multiple scatterings in the hadronic absorber.

True associations between the FVTX and MuTr are 
determined after a statistical subtraction using MuTr+FVTX combinations 
from different events, which account for the wrong associations. The 
background-subtraction procedure is described in Sect. III.A. The 
tracks are separated into (i) hadron-rich and (ii) muon-rich samples.  
The elliptic flow for each of these samples is determined by measuring 
the angle between tracks as described in Section III.B.  Section IV.A 
describes the analysis procedure. Simulation studies are used to 
extract v2 of charged hadrons and inclusive muons from the hadron- 
and muon-rich samples. The heavy-flavor-decay 
contribution to the muon sample is then obtained after simulating the 
light hadron decays into muons.

\subsection{Event and track selection}

The large amount of absorbing material between the FVTX and MuTr is 
effective at stopping hadrons, but it also causes significant issues 
with matching tracks between the two detectors. Particles can undergo 
multiple scatterings in the absorbing material causing a single MuTr 
track to be matched with multiple FVTX tracks. A ``mixed-event" 
subtraction technique is employed to precisely estimate the substantial 
background from mismatched tracks and multiple scatterings.

\begin{figure} [hb]
\includegraphics[width=1.0\linewidth]{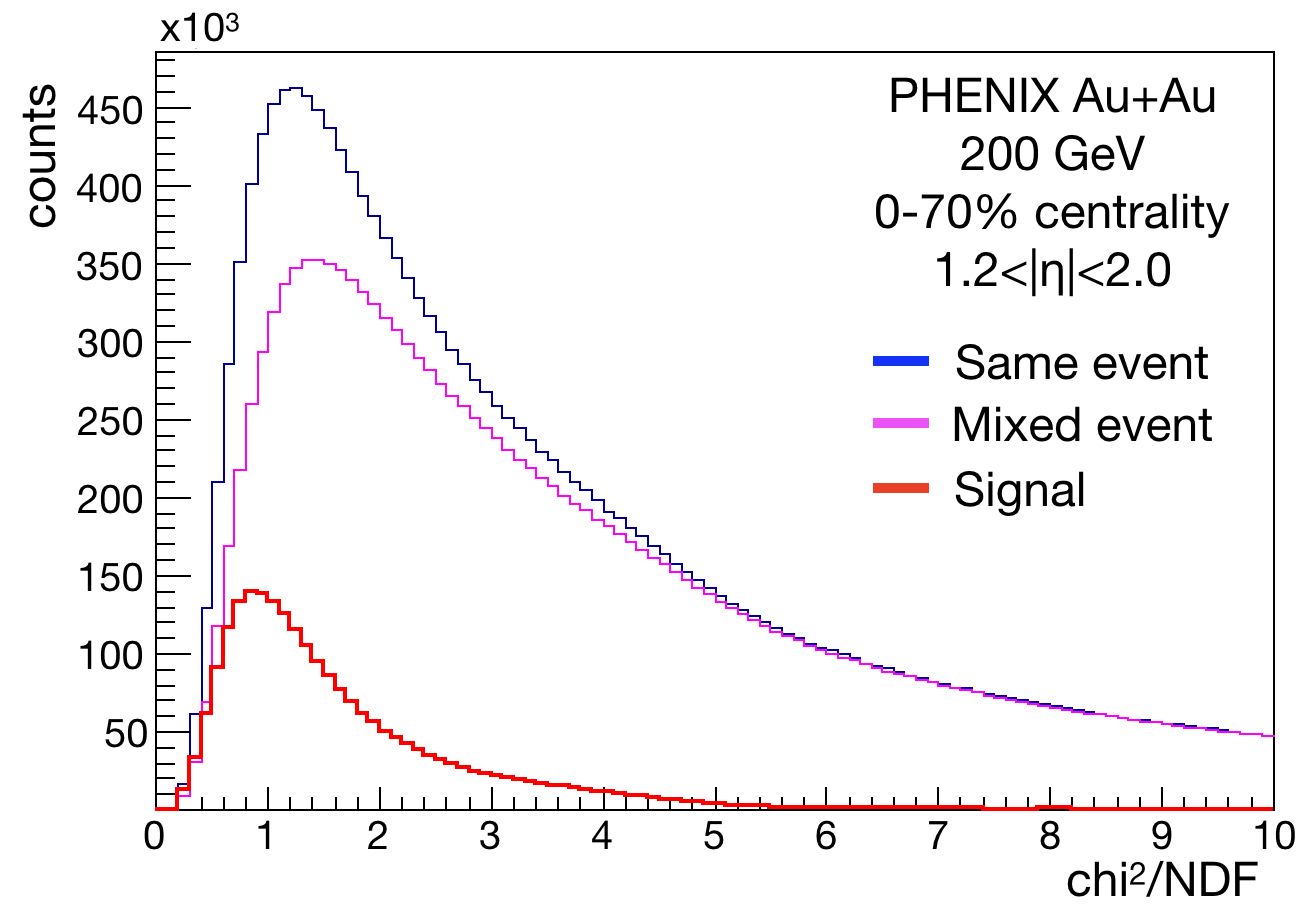}
\caption{The $\chi^{2}$-FVTXMuTr distributions for the same- and 
mixed-event matchings in the range $1<p_T<1.25$ GeV/$c$ with the true 
matches extracted from the fit.}
\label{fig:matching}
\end{figure}

We quantify the quality of MuTr-FVTX track matching with the 
$\chi^2$-FVTXMuTr variable, which includes the residuals between the 
track projections in three different Z planes between FVTX and MuTr. 
The $\chi^2$-FVTXMuTr distribution of same-event matching include true 
associations, where the MuTr track and the FVTX track correspond to the 
same particle, and ``fake" matchings, for which the MuTr and FVTX tracks 
are from different particles. The $\chi^2$-FVTXMuTr distribution of 
mixed-event associations include only fake matchings.  Mixed-event 
associations are constructed from events that have a similar Z-vertex 
and event centrality. The same-event $\chi^2$-FVTXMuTr distribution is 
described by the function:$f(\chi^2) = f_{\rm true}(\chi^2) + 
f_{\rm fake}(\chi^2)$, where $f_{\rm true}$ and $f_{\rm fake}$ 
represent the true and fake associations, respectively. The line shape 
of $f_{\rm fake}$ is obtained from mixed-event matchings . A Landau 
distribution best represents $f_{\rm true}$, regardless of the $p_T$ of 
the muon and the event centrality. The function $f(\chi^2)$ is fit to 
the same-event $\chi^2$-FVTXMuTr distribution as shown in 
Fig.~\ref{fig:matching}. The integral of $f_{\rm true}$ corresponds to 
the yield of particles measured by the MuTr and FVTX coming from the 
vertex.

Loose track-selection criteria are used for initial data production. 
Then, the selections detailed in Table~\ref{table:track_cuts} are 
applied to maximize the purity of heavy-flavor muons in the data set. 
These selections were determined using full {\sc geant4} simulations 
embedded in real data. The embedded simulations were also used to 
validate the background subtraction procedure, where it is important 
that the background corresponds to real data.

\begin{table}[htb]
\caption{\label{table:track_cuts}
Event- and track-selection criteria.}
\begin{ruledtabular} \begin{tabular}{ccc}
Variable & south arm & north arm \\
\hline
Centrality & 0\%--70\% & 0\%--70\% \\
$|{\rm BBC}_z|$ & $<$10 cm & $<$10 cm \\
$|{\rm DCA}_R|$ & $<$0.05 cm & $<$0.05 cm \\
$\chi^{2}$-FVTXMuTr & $<3$ & $<3$ \\
pDG0 & $<$60 mm$\cdot$ GeV/$c$ & $<$35 mm$\cdot$ GeV/$c$ \\
pDDG0 & $<$40 mrad$\cdot$ GeV/$c$ & $<$40 mrad$\cdot$ GeV/$c$ \\
MuTr$\chi^{2}$ & $<5$ & $<5$ \\
MuTrhits & ${\ge}14$ & ${\ge}16$ \\
MuIDhits & ${\ge}4$ & ${\ge}4$ \\
MuID$\chi^{2}$ & $<5$ & $<5$ \\
$|pz|$ & $>$3 GeV/$c$ & $>$3.6 GeV/$c$ \\
lastgap & = 2, 3, or 4 & = 2, 3, or 4 \\
\end{tabular} \end{ruledtabular} 
\end{table}

The variables for the event- and track-selection criteria that are 
tabulated in Table~\ref{table:track_cuts} are:

\begin{description}

\item[BBC$_z$] The $z$ vertex position of the initial collision point 
as determined by the BBCs

\item[DCA$_R$] Radial distance of closest approach (i.e., the measure 
of the distance from initial collision point at which the particle is 
produced)

\item[$\chi^{2}$-FVTXMuTr] Quality of track matching between FVTX and MuTr

\item[DG0] Distance between MuID 1st hit and extrapolated track

\item[DDG0] Angular distance between MuID 1st hit and extrapolated track

\item[MuTr$\chi^2$] Quality of matching between reconstructed track
projection and hits in the MuTr

\item[MuTrhits] Number of MuTr hits for particle

\item[MuIDhits] Number of MuID hits for particle

\item[MuID$\chi^2$] Quality of matching between reconstructed track
projection and hits in the MuID

\item[$pz$] Particle $z$ momentum

\item[lastgap] Last hit position in MuID

\end{description}

\begin{figure}
\includegraphics[width=1.0\linewidth]{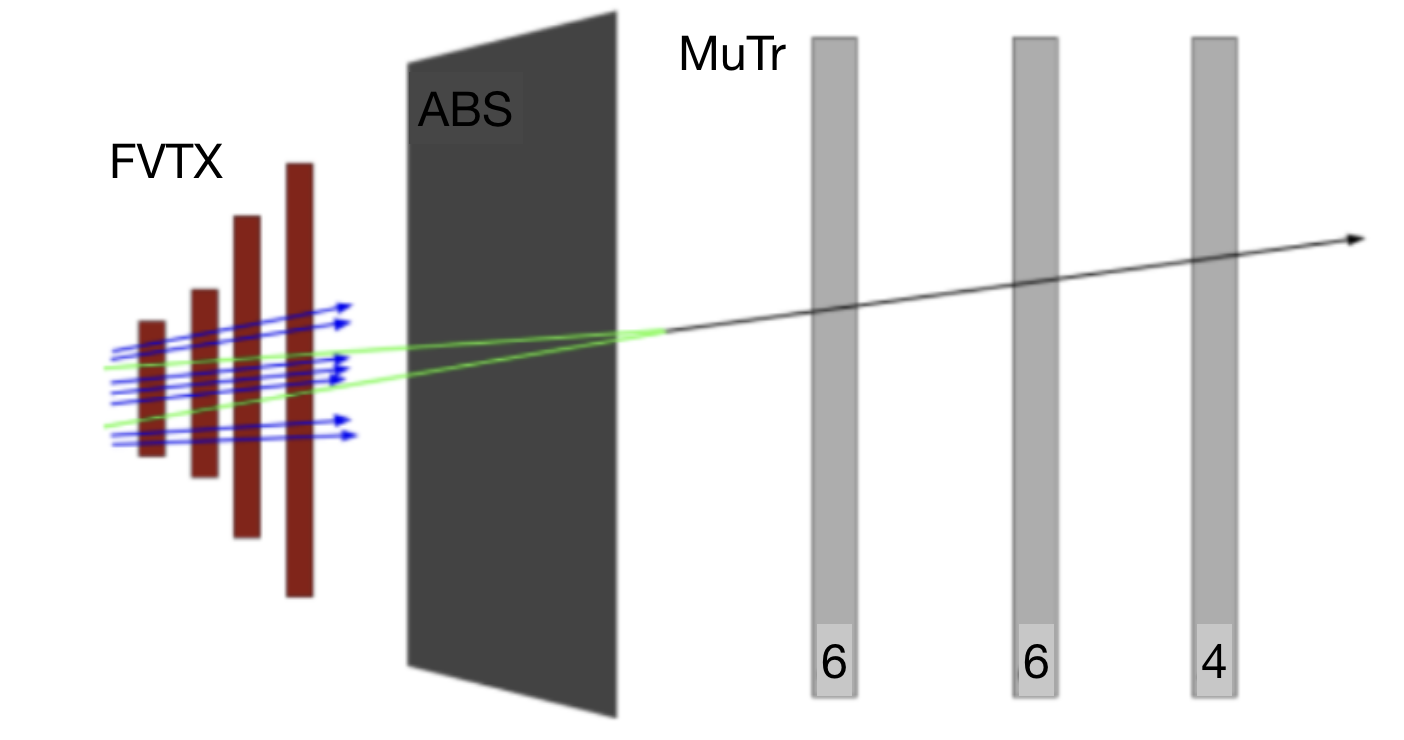}
\caption{Illustration of a single MuTr track being matched with all 
FVTX tracks within $2\sigma$ of the residual $\eta$ and $\phi$ 
distribution determined by the FVTX and MuTr.}
\label{fig:track_matching}
\end{figure}

\subsection{Event-plane method and azimuthal-anisotropy measurements}

In an off-center collision of two nuclei the reaction plane is defined 
as the plane formed by the impact parameter vector and the initial 
collision axis for each event. In practice, the reaction plane is not 
directly observable, but the azimuthal distribution of the particles 
detected in the final state can be used to determine the EP that 
contains both the beam direction and the azimuthal direction of maximum 
particle density. The event-flow vector $Q_n$ and the EP angle $\Psi_n$ 
for each $n^{th}$ harmonic are defined by the following:

\begin{align}
    Q_x&=\sum_i\cos(n\phi_i)\quad\quad 
    Q_y&=\sum_i\sin(n\phi_i)
\end{align}

\begin{equation}
    \Psi_n = atan(Q_y/Q_x),
\end{equation}

\noindent where $\phi_i$ are the azimuthal angles of the particles used 
for EP determination.

The $v_n$ observed with respect to the event plane is given by:

\begin{equation}
    v_n^{\rm obs}=\langle \cos[n(\phi-\Psi_n)]\rangle, 
\end{equation}

\noindent where $\langle\,\rangle$ denotes an average over all particles 
and all selected collision events. The true flow is obtained by 
dividing the observed flow by the corresponding EP resolution:

\begin{equation}
    v_n=\frac{v_n^{\rm obs}}{\mathcal{R}_n}
\end{equation}

\begin{figure}
\includegraphics[width=1.0\linewidth]{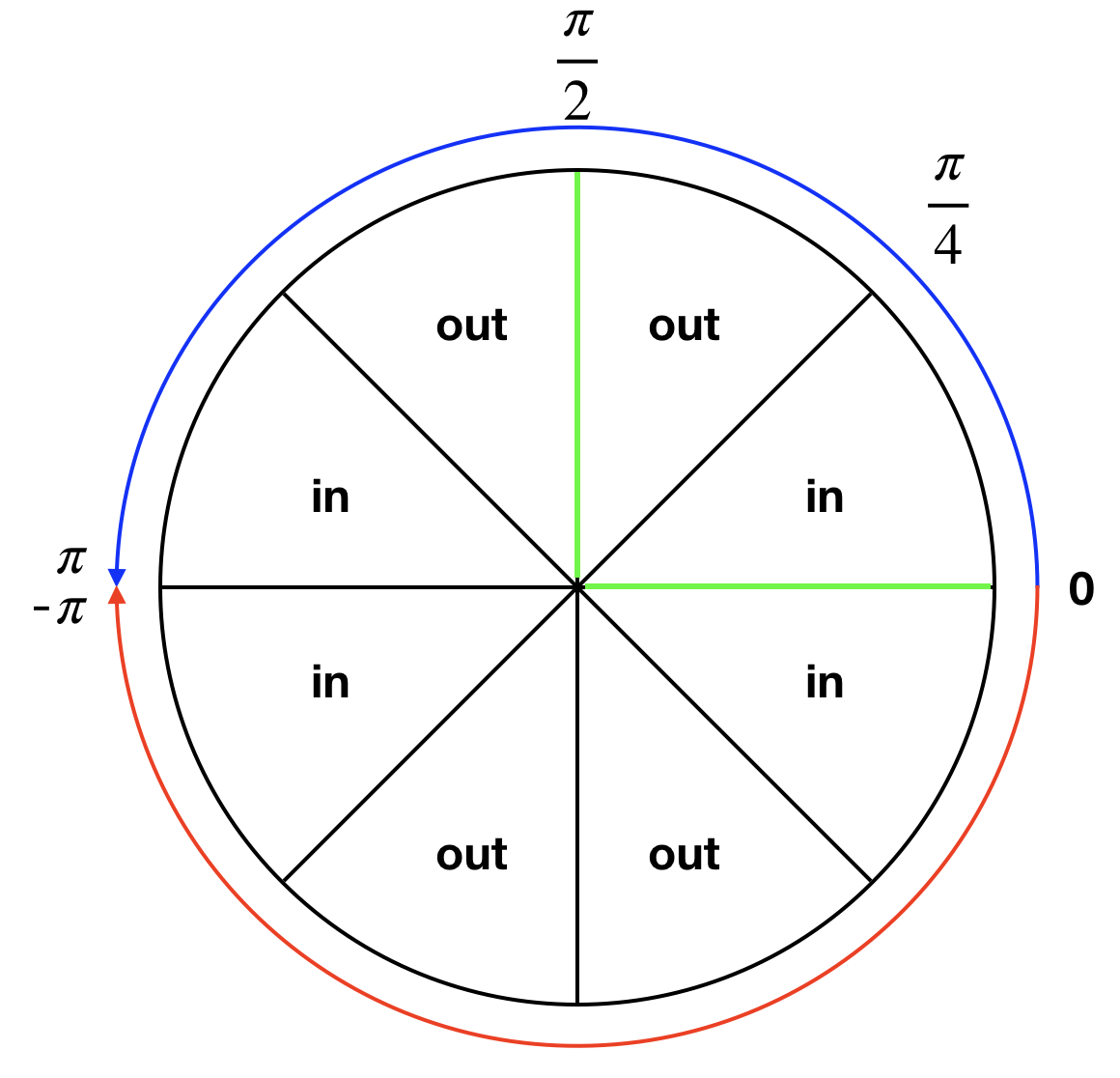}
\caption{Diagram of in plane and out of plane particle emission 
relative to the EP angle.}
\label{fig:InOut}
\end{figure}

The second-order EP angle, $\Psi_2$, is measured using the FVTX 
detectors on the north and south sides of the central-arm 
spectrometers. To avoid short-range correlations that are not 
associated with the collective flow, it is critical to have a large 
rapidity gap between the detector used to determine the event plane and 
the detector used in the flow measurement. Thus, particles detected in 
the north muon arm are correlated with the event plane measured in the 
south FVTX, and vice versa, corresponding to a minimal $\eta$ gap of 
2.4. The event-plane resolution is estimated using the three subevent 
method~\cite{Poskanzer:1998yz},

\begin{equation}
    \mathcal{R}_2^{a}=\sqrt{\frac{\langle \cos(2(\Psi_a-\Psi_b))\rangle \langle \cos(2(\Psi_a-\Psi_c))\rangle}{\langle \cos(2(\Psi_b-\Psi_c))\rangle}}
\end{equation}

\noindent where $\Psi_a$ is determined from the primary FVTX arm, 
$\Psi_b$ is determined from the opposite side FVTX arm, and $\Psi_c$ is 
determined from the CNT detector. Following the determination of the EP 
angle and associated resolution, the azimuthal anisotropy is measured 
by counting the number of particles in different $\Delta \phi$ bins 
relative to the event-plane angle. By distinguishing the azimuthal 
angle of emission for each individual particle as either ``in-plane" or 
``out-of-plane" relative to the event plane angle 
(Fig.~\ref{fig:InOut}) the azimuthal anisotropy is determined for a 
given transverse-momentum bin by:

\begin{equation}
    v_2^{\rm obs}=\frac{\pi}{4}\frac{N_{\rm in}-N_{\rm out}}{N_{\rm in}+N_{\rm out}}
\end{equation}

The statistical uncertainties $\sigma_{\rm in}$ and $\sigma_{\rm out}$ 
in the number of particles detected in-plane and out-of-plane ($N_{\rm 
in}$ and $N_{\rm out}$) are propagated to the uncertainty 
$\sigma_{v^{\rm obs}_{2}}$ in the $v_2^{\rm obs}$ measurement.

\begin{equation}
    \sigma_{v^{\rm obs}_{2}}=\frac{\pi}{2(N_{\rm in}+N_{\rm out})^2}\cdot\sqrt{(N_{\rm out}\sigma_{\rm in})^2+(N_{\rm in}\sigma_{\rm out})^2}
\end{equation}

\begin{figure}
\includegraphics[width=1.0\linewidth]{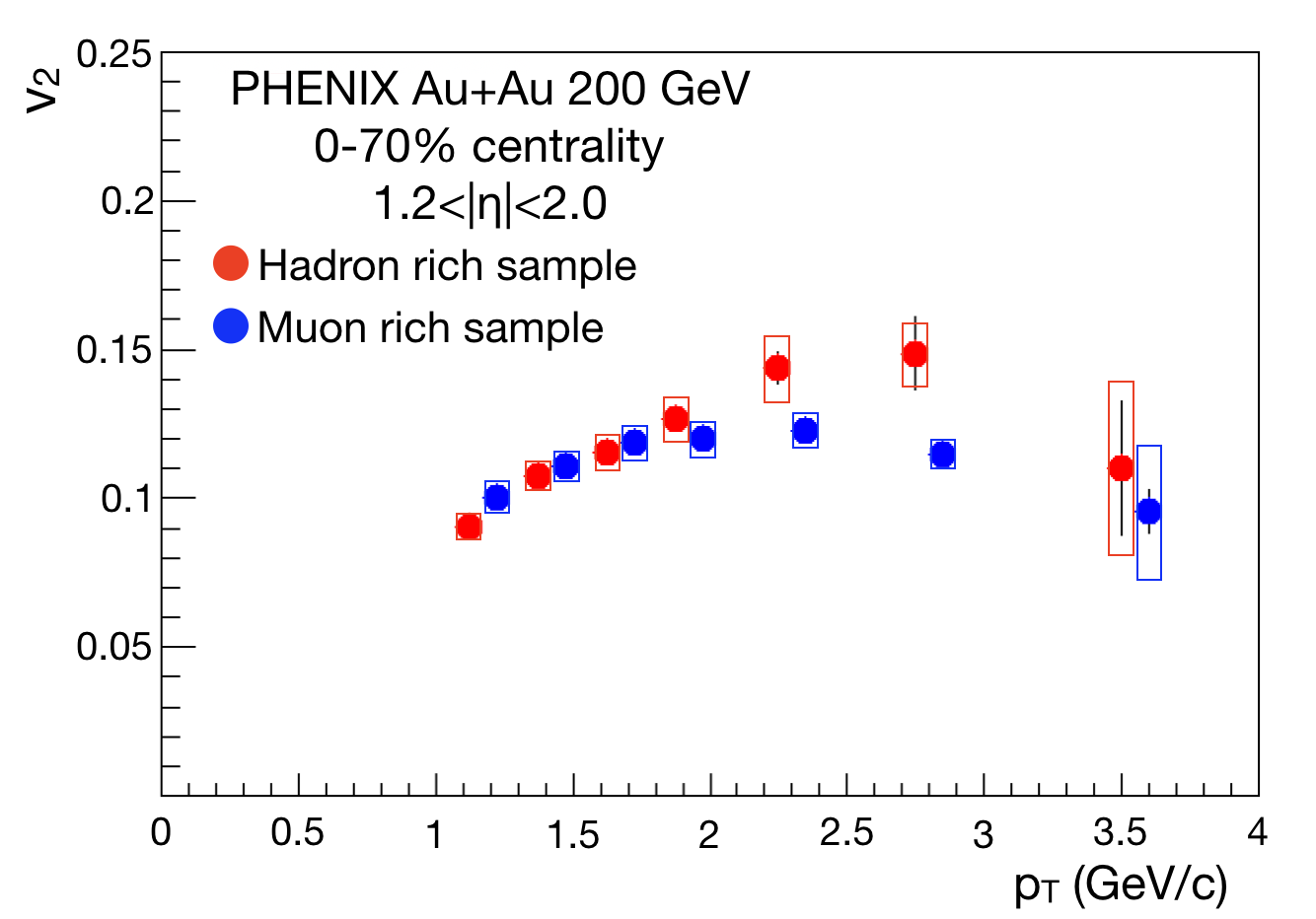}
\caption{Elliptic flow of charged particles in hadron-rich and 
muon-rich samples. For clarity the data for the muon-rich sample are 
offset by 0.1 GeV/$c$ to higher $p_T$.}
\label{fig:HRich_v2}
\end{figure}

 \section{Analysis Procedure}
\label{sec:Procedure}

\subsection{Simulation studies}

We separate the particles into two distinct samples: ({\it i}) the 
light-hadron-rich sample comprises particles that stop in layers two 
and three of the MuID; and ({\it ii}) the muon-rich sample comprising 
particles which can penetrate layers three and four of the MuID. Using 
the method described in Section~\ref{sec:Methods} we measure $v_2$ of 
both samples as shown in Fig.~\ref{fig:HRich_v2}.

\begin{figure}[htbp]
\includegraphics[width=1.0\linewidth]{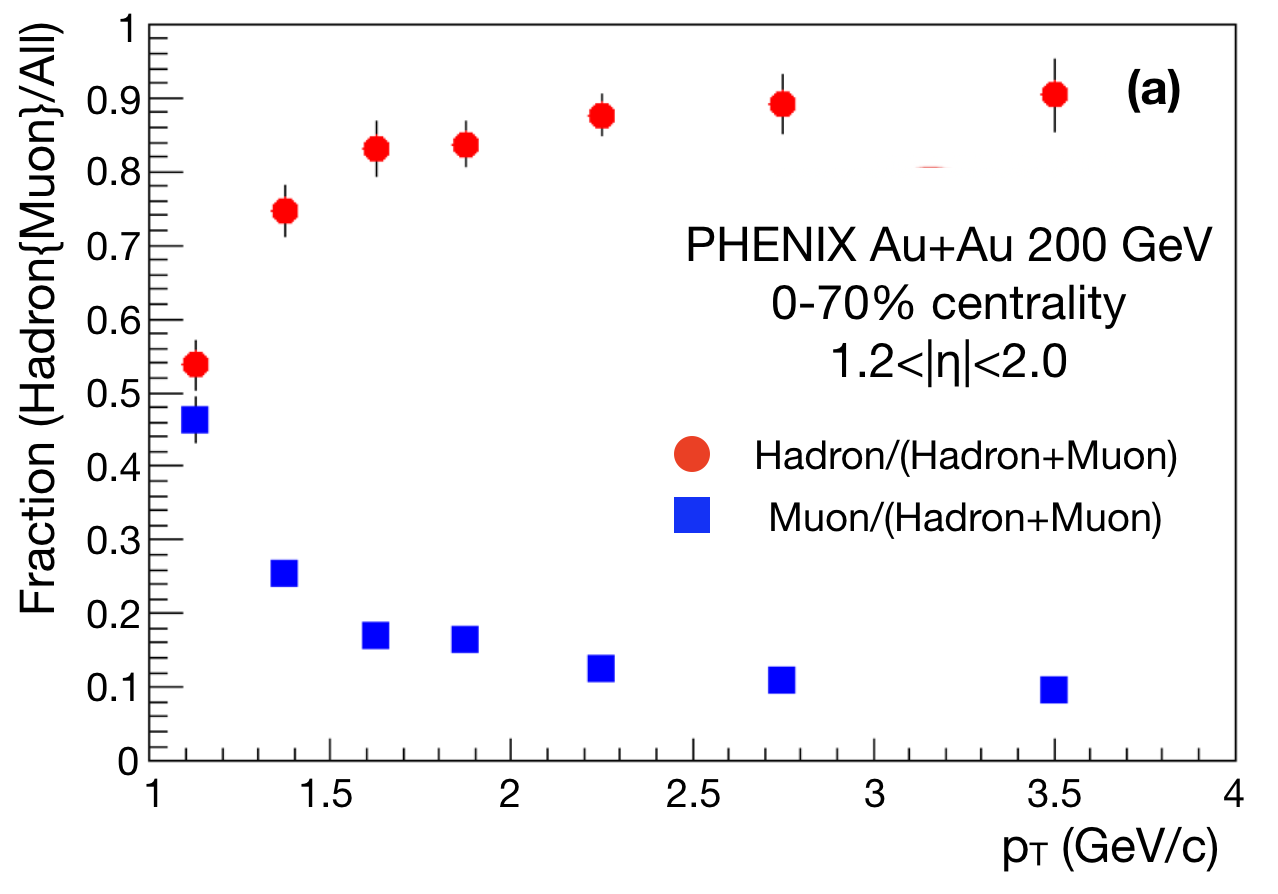}
\includegraphics[width=1.0\linewidth]{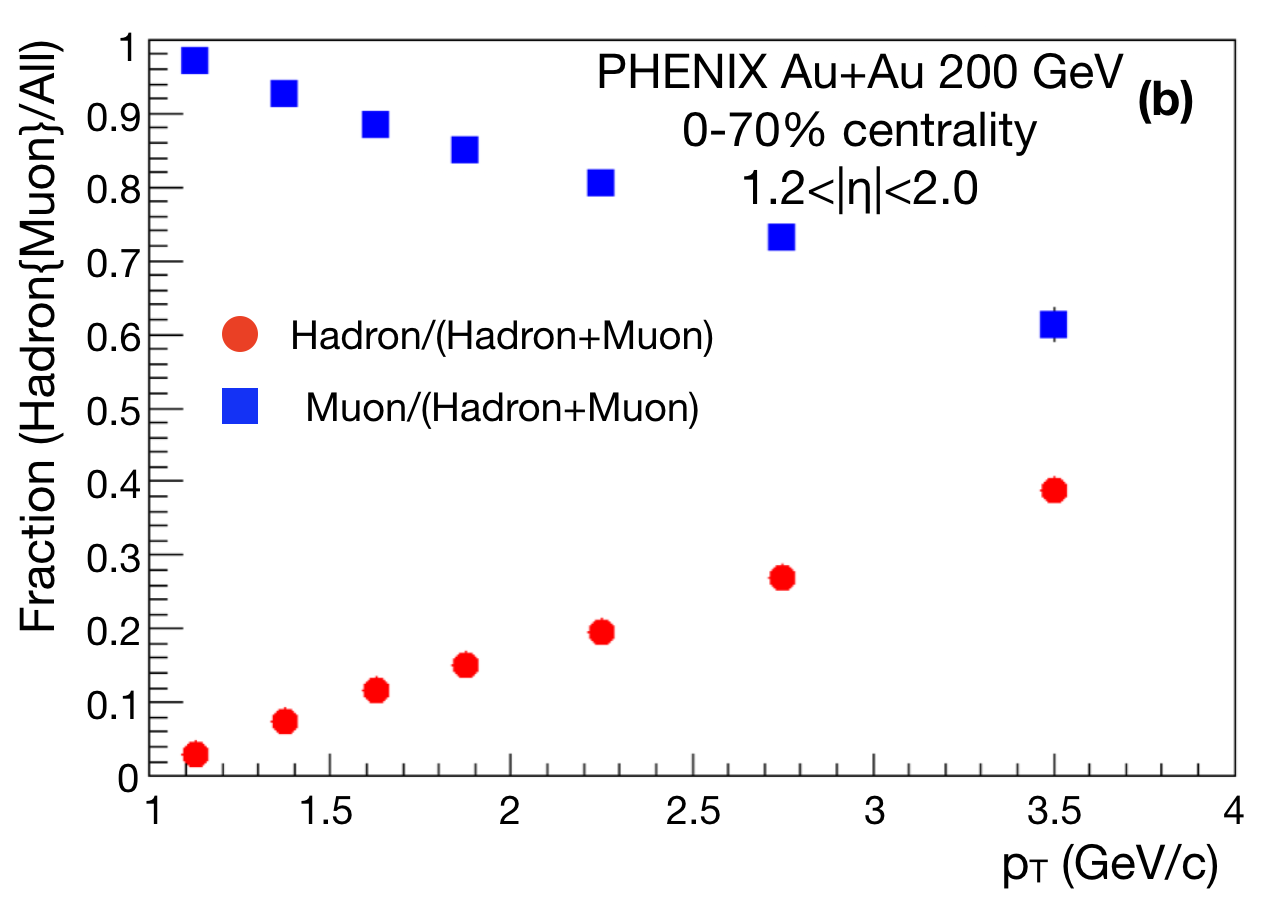}
\caption{Simulated hadron and muon particle fractions as a function of 
$p_T$ for (a) particles that stop in layers two and three of the MuID 
(hadron-rich sample), and (b) particles that penetrate layers three and 
four of the MuID (muon-rich sample).
\label{fig:muon_rich_ratio}}
\end{figure}

A full {\sc pythia3} + 
{\sc geant4}~\cite{Bierlich:2022pfr,GEANT4:2002zbu} detector simulation 
was employed to determine the particle composition of both samples. 
This simulation was validated to reproduce the $\phi$ and $\eta$ 
distributions seen in the data, and remaining discrepancies are resolved 
using fiducial cuts. Detectable particles are generated by {\sc 
pythia3}, they include kaons, pions, protons, and muons. The K/$\pi$, 
$K^+/K^-$, and $\pi^+/\pi^-$ yield ratios provided by {\sc pythia3} are 
modified according to extrapolations based on measurements performed at 
midrapidity by PHENIX~\cite{PHENIX:2013kod}, and at forward rapidity by 
BRAHMS~\cite{BRAHMS:2016klg} in Au$+$Au collisions at 
$\sqrt{s_{_{NN}}}=$200 GeV. The charged particle $p_T$ distribution is 
also weighted according to PHENIX and BRAHMS data. {\sc pythia3} events 
containing these detectable particles are used as input to the {\sc 
geant4}-based detector simulation. Approximately 10$^{-4}$ of the light 
hadrons can penetrate the first muon-spectrometer absorber. Only events 
where at least one of the particles produces more than two hits in the 
MuTr detector volume are kept. Fewer than 2\% of the hadrons producing 
hits in MuTr are estimated to be protons, thus they were excluded in 
this analysis. The {\sc geant} hits are then merged with hits from a 
minimum-bias real-data event and reconstructed by the tracking 
algorithms. Figure~\ref{fig:muon_rich_ratio} shows the composition of 
light hadrons and muons in samples ({\it i}) and ({\it ii}) as a 
function of reconstructed track $p_T$ as obtained from the full {\sc 
pythia3}+{\sc geant4} simulation.

The $v_2$ of hadron-rich ($v_2^{\rm h-rich}$) and muon-rich
($v_2^{\rm \mu-rich}$) samples comprise the following terms:

\begin{equation}
\label{eq:v2_composition}
    v_2^{h-{\rm rich}} = \frac{N_h^{h-{\rm rich}}}{N^{h-{\rm rich}}} v_2^{h} + \frac{N_{\mu}^{h-{\rm rich}}}{N^{h-{\rm rich}}} v_2^{\mu} \\
\end{equation}

\begin{equation}
\label{eq:v2_comp_2}
    v_2^{\mu-{\rm rich}} = \frac{N_h^{\mu-{\rm rich}}}{N^{\mu-{\rm rich}}}v_2^{h} + \frac{N_{\mu}^{\mu-{\rm rich}}}{N^{\mu-{\rm rich}}} v_2^{\mu},
\end{equation}

\noindent where the fractions are the light-hadron and inclusive muon 
compositions of both samples as shown in 
Fig.~\ref{fig:muon_rich_ratio}. 
In these equations $N_h^{h-{\rm rich}}$ and $N_\mu^{h-{\rm rich}}$ are the
numbers of hadrons and muons in the hadron-rich region. Similarly,
$N_h^{\mu-{\rm rich}}$ and $N_\mu^{\mu-{\rm rich}}$ are the hadron and muon
counts in the muon-rich region.
The total counts for hadrons and muons are $N^{h-{\rm rich}}$ 
and $N^{\mu-{\rm rich}}$.  The light-hadron and muon-elliptic-flow 
values $v_2^{h}$ and $v_2^{\mu}$ are obtained by solving the set of 
Eqs.~\ref{eq:v2_composition}~and~\ref{eq:v2_comp_2}.

\begin{figure}[ht]
\centering
\includegraphics[width=1.0\linewidth]{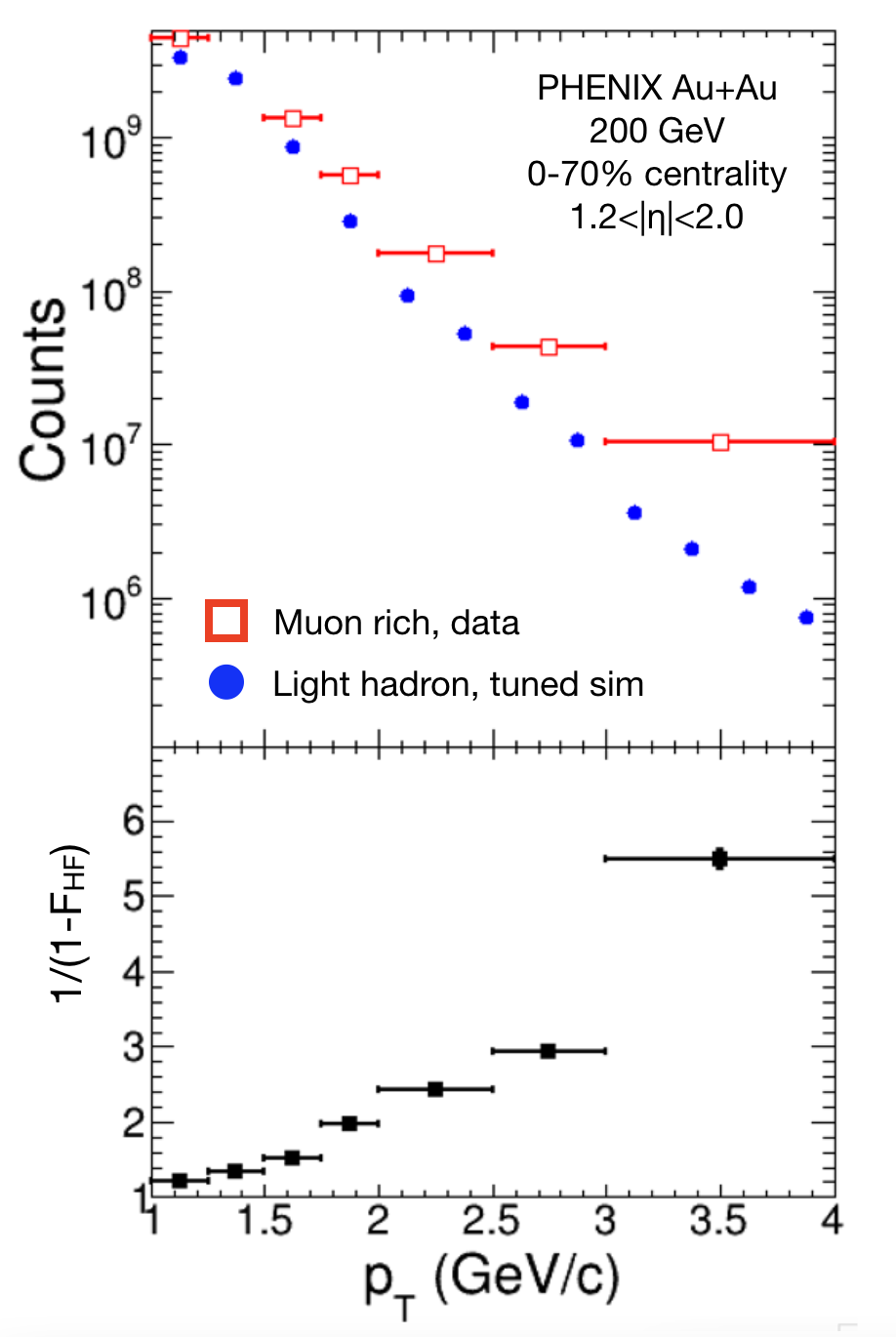}
\caption{The muon-rich data compared to muon-rich simulation 
(with heavy-flavor contribution excluded) and the ratio between 
the two as a function of $p_T$.}
\label{fig:F_HF}
\end{figure}

The muon sample comprises light-hadron decays and heavy-flavor decays. 
Therefore, its flow can be determined as:

\begin{align}
v_2^{\mu} &= F_{\rm HF} v_2^{\rm HF} + \left(1-F_{\rm HF}\right)v_2^{h\to\mu} \\
F_{\rm HF} &= \frac{N_{HF\to\mu}}{N_{\rm LH\to\mu}+N_{\rm HF\to\mu}},
       \label{eq:muon_v2_composition}
\end{align}

\noindent where $F_{\rm HF}$ is the heavy-flavor-decay contribution to 
the muon sample, and $v_2^{h\to\mu}$ is the elliptic flow of muons from 
light-hadron decays. The light-hadron contribution $N_{\rm LH\to\mu}$ 
to the muon-rich sample is obtained from the detector simulation using 
as input the measured counts in the light-hadron sample 
$N^{h-{\rm rich}}$. The light-hadron projection to the muon-rich region 
and the ratio to the muon-rich data are shown in Fig.~\ref{fig:F_HF}.  
The elliptic flow $v_2^{h\to\mu}$ of muons from light-hadron decays 
seen in the muon-rich sample is also determined from the detector 
simulation. The phi distribution of the generated light flavor hadrons, 
which serves as input to the detector simulation, is weighted by the 
measured $v_2^h$ extracted from Eqs. 9 and 10. The decay of these 
simulated hadrons is processed in the detector simulation and the 
elliptic flow of the daughter muons is determined by comparing the phi 
distribution of the weighted parent hadrons and the produced daughter 
muons.

\subsection{Systematic uncertainties for $v_2$}

Four main sources of systematic uncertainties were considered: 
false-match determination, track selections, particle composition, and 
event-plane determination.

\subsubsection{Background determination} 

We evaluate the uncertainty that the false-match determination method 
introduces to the measurement of $v_2$. In Section~\ref{sec:Methods} 
the true matches are obtained after determining a $p_T$-dependent 
normalization scale for the mixed-event sample. This method assumes 
that the range where the $\chi^2$-FVTXMuTr distribution is fitted to 
obtain the normalization is dominated by false matches in the 
same-event sample. This allows for a direct comparison to the 
mixed-event sample, which is comprised entirely of false matches. To 
account for the uncertainties associated with this assumption, the 
normalization factor was increased and decreased by two standard 
deviations of the value returned by the fit.  Then, the remainder of 
the analysis was performed identically. Any difference in the final 
$v_2$ value is due to the false-match determination method.

\subsubsection{Track selections} 

There are numerous track selections that are used to improve the signal 
over background in the dataset and to exclude misreconstructed 
particles. To determine the influence of these selections on the final 
$v_2$ values, the analysis was repeatedly performed by independently 
varying each cut by $10\%$. The cuts that were varied are pDG0, pDDG0, 
MuTr$\chi^2$, MuTrhits, MuIDhits, and MuID$\chi^2$. For this study we 
did not vary $\chi^2$-FVTXMuTr because uncertainties associated with 
this parameter are already considered in our studies on the background 
determination. Because these cuts are all correlated, the largest 
difference for each $p_T$ bin was taken as the associated systematic 
uncertainty.
 
\subsubsection{Particle composition} 

Another source of systematic uncertainty is associated with the 
particle composition in the simulations described in 
Sec.~\ref{sec:Procedure}. The particle composition was tuned to match 
previously published PHENIX~\cite{PHENIX:2013kod} results from Au$+$Au 
collisions for kaon and pion ratios, as well as charged hadron spectra.  
However, these results are from midrapidity ($|\eta|<0.35$) 
measurements, unlike our analysis. The $p_T$-integrated 
forward-rapidity measurements from BRAHMS~\cite{BRAHMS:2016klg} 
indicated no strong rapidity dependence for kaon and pion ratios.  As a 
systematic check, we varied the particle ratios in the simulation 
within the uncertainties of the BRAHMS results and remeasured the 
$v_2$.  Additional uncertainties arise from using the tuned simulation 
to decompose the $v_2$ measurements into hadron and muon contributions 
and counting the number of hadrons and muons in any given $p_T$ bin.  
These two uncertainties are added in quadrature to obtain the total 
systematic uncertainty from particle composition.

\subsubsection{Event-plane determination} 

In the standard analysis we use the FVTX to determine the EP angle and 
the muon arm opposite to the event-plane detector to measure $v_2$. To 
evaluate how various detector inefficiencies and differences in 
rapidity gaps between the muon detectors and the detectors used to 
determine the EP angle influence the final $v_2$ results, a different 
set of detectors was used to perform the measurement. Namely, the 
analysis was performed again using the central-arm tracking detectors 
($|\eta|<0.35$) as the event-plane detector for both north and south 
muon-arm measurements.

\subsubsection{Total systematic uncertainty} 

The relative systematic uncertainty from each source is listed in 
Table~\ref{Table:systematic}. Independent evaluations for each of the 
sources were added in quadrature to obtain the total systematic 
uncertainty of the $v_2$ measurements.

\begin{table}[ht]
\caption{\label{Table:systematic}
Relative systematic uncertainty in percent for the measurements of
$v_2$ for hadrons and heavy-flavor muons.}
\begin{ruledtabular} \begin{tabular}{ccccc}
$p_T$[GeV/$c$] & Background & Composition & Track cuts & EP\\
\hline
1.00--1.25 & 1.0 & 4.4 & 4.7 & 1.0    \\
1.25--1.50 & 0.9 & 4.5 & 3.5 & 0.5   \\
1.50--1.75 & 0.6 & 4.6 & 4.3 & 0.9   \\
1.75--2.00 & 0.6 & 4.3 & 6.0 & 4.3   \\
2.00--2.50 & 0.2 & 4.1 & 4.0 & 1.3   \\
2.50--3.00 & 0.4 & 5.6 & 4.3 & 14    \\
3.00--4.00 & 13  & 6.3 & 20  & 8.0   \\
\end{tabular} \end{ruledtabular}
\end{table}

\section{Results and Discussion}
\label{sec:Results}

\begin{figure}[ht]
\includegraphics[width=1.0\linewidth]{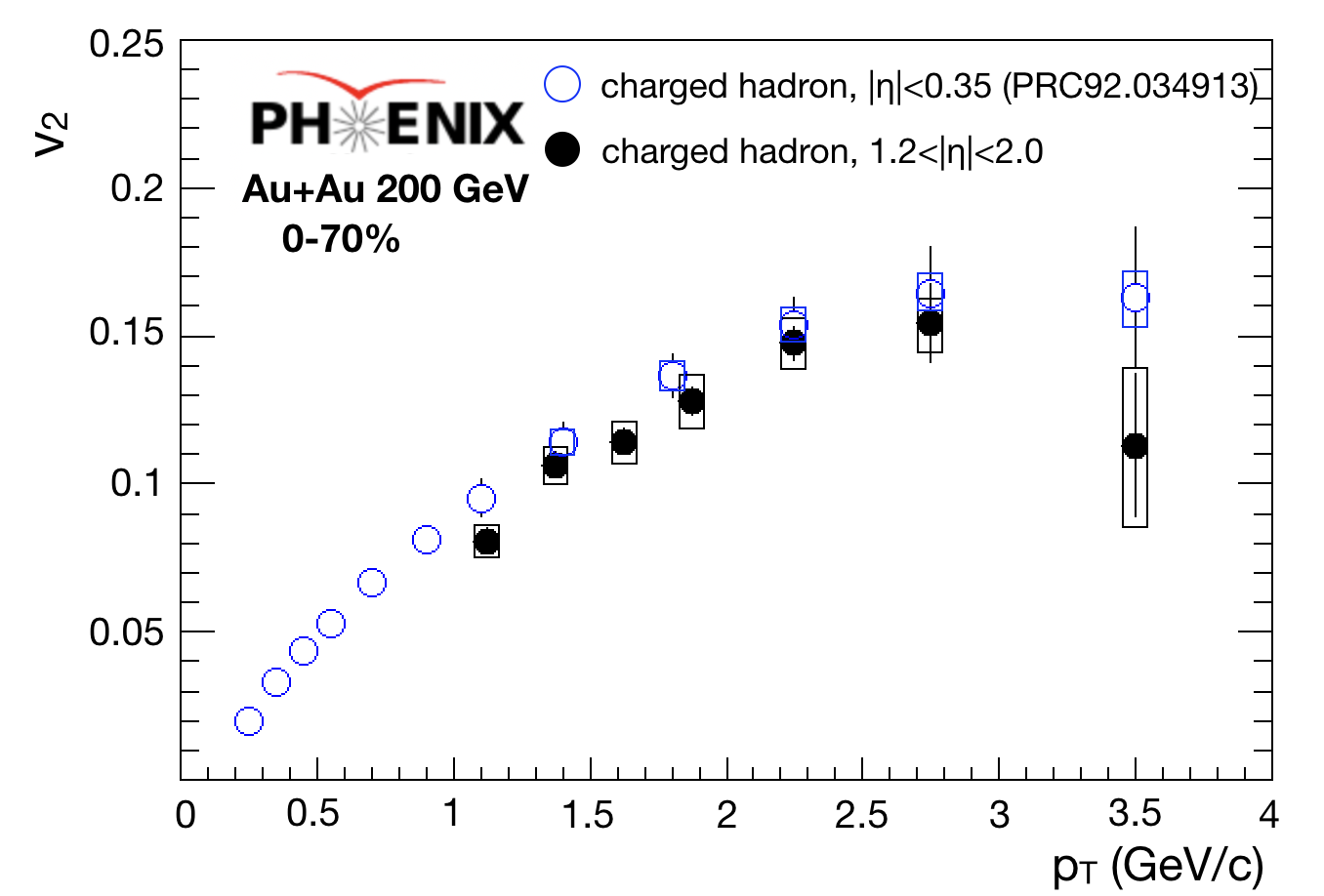}
\caption{The value of $v_2(p_T)$ of charged hadrons at forward rapidity 
compared to previous midrapidity results. The error bars show 
statistical uncertainty and the boxes show systematic uncertainties}
\label{fig:hadron_v2}
\end{figure}

The $v_2$ measurements in the PHENIX muon arms for charged hadrons and 
open-heavy-flavor muons are presented here as a function of $p_T$ 
($1.2<|\eta|<2.0$). The results are compared to previous PHENIX 
measurements~\cite{PHENIX:2014yml} at midrapidity ($|\eta|<0.35$). 
Figure~\ref{fig:hadron_v2} shows the measurements of $v_2(p_T)$ for 
charged hadrons. The values of $v_2(p_T)$ at forward rapidity are 
systematically below those at midrapidity by $\approx$10\%. This is 
comparable to the size of systematic uncertainties, which are 
point-to-point correlated. The PHOBOS results on the rapidity 
dependence of $p_T$-integrated-charged-hadron 
$v_2$~\cite{PHOBOS:2004vcu} show a similar decrease in $v_2$ between 
the rapidity ranges of the PHENIX measurements. A caveat to the 
$v_2(p_T)$ comparison is that our forward rapidity measurement does not 
include protons and antiprotons that are included in the charged-hadron 
measurements at midrapidity. In the higher-transverse-momentum range 
($p_{T} > 2$ GeV/$c$) at midrapidity there will be a proton 
contribution to the charged-hadron $v_2$ as the proton/pion ratio 
increases in this range~\cite{PHENIX:2003iij}.

\begin{figure}[ht]
\includegraphics[width=1.0\linewidth]{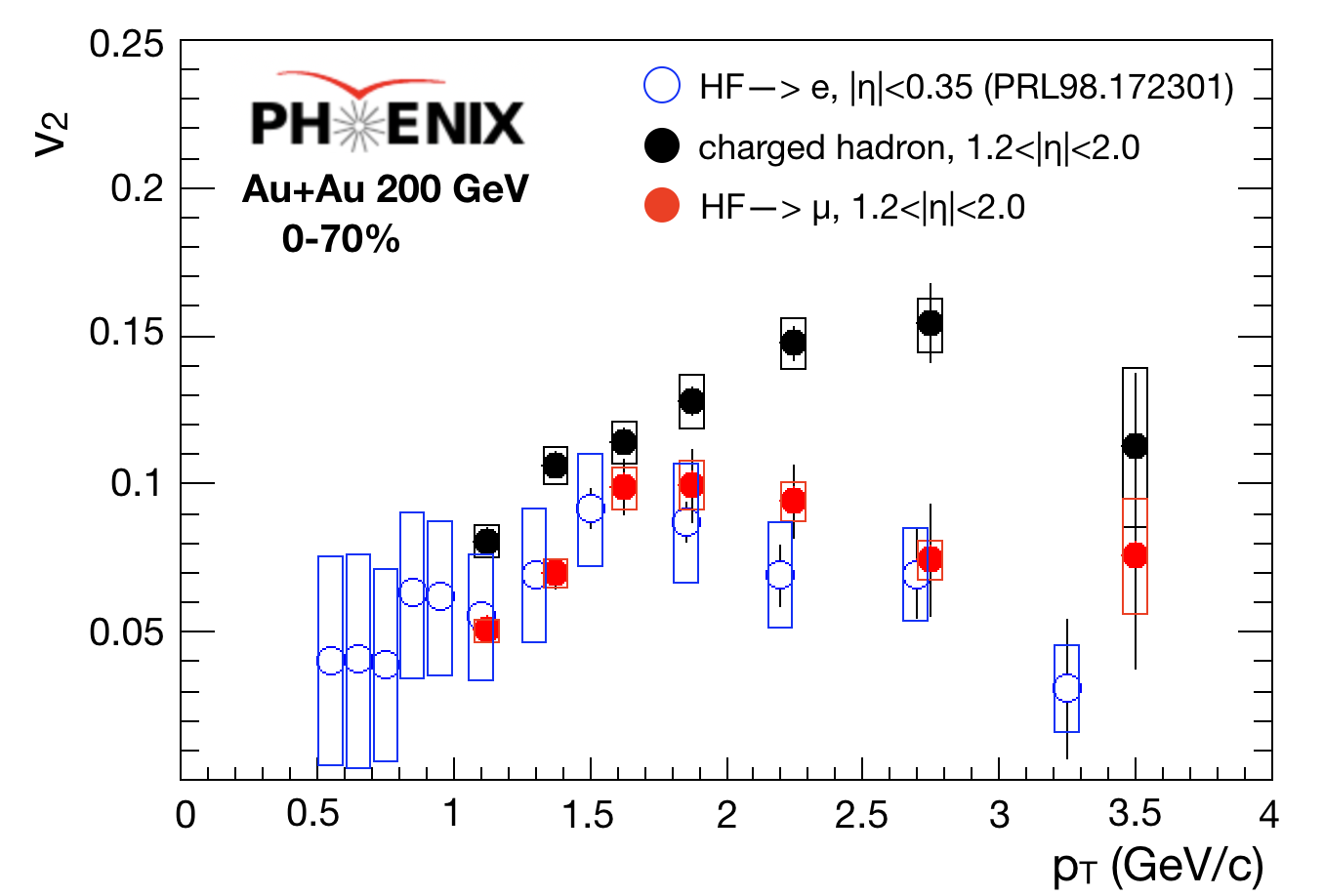}
\caption{The value of $v_2(p_T)$ of heavy-flavor muons and charged 
hadrons at forward rapidity and heavy-flavor electrons at midrapidity.}
\label{fig:HF_v2}
\end{figure}

Figure~\ref{fig:HF_v2} shows the $v_2$ of open-heavy-flavor muons at 
forward rapidity compared to charged hadrons in the same rapidity range 
and previous PHENIX results of open-heavy-flavor electrons at 
midrapidity~\cite{PHENIX:2006iih}. The heavy-flavor muons exhibit 
smaller $v_2$ than the light hadrons, which is to be expected given the 
mass ordering of particle interactions with the QGP. Similar to the 
observations in the charged-hadron measurement, there is no pronounced 
rapidity dependence for open-heavy-flavor flow. There appears to be a 
slight shift to higher $p_T$ in the results for heavy-flavor muons as 
compared to heavy-flavor electrons. This may be due to the large mass 
difference between the decay leptons.  Despite this apparent shift, the 
heavy-flavor electron and muon $v_{2}(p_{T}$) have similar magnitude 
indicating no clear difference in heavy-quark interactions with the QGP 
in the two measured rapidity ranges.

\section{Summary}
\label{sec:Conclusion}

We have measured the elliptic anisotropy, $v_2$, of charged hadrons and 
open-heavy-flavor muons in Au$+$Au collisions at 
$\sqrt{s_{_{NN}}}=200$~GeV at forward rapidity ($1.2<|\eta|<2.0$). The 
measurements were performed using data sets collected in 2014 and 2016 
corresponding to an integrated luminosity of 14.5~nb$^{-1}$. This is 
the largest Au$+$Au data set collected by PHENIX. Measuring 
${\rm DCA}_R$ with the PHENIX (F)VTX detectors and we use an FVTX-MuTr 
matching method to successfully overcome the large background in the 
heavy-flavor-muon sample originating from light-hadron decays and track 
mismatching before and after the hadron absorber. Using samples 
detected in different regions of the muon-identification detectors, the 
light hadrons were separated from inclusive muons, and the 
heavy-flavor-muon contributions were extracted.

The values of $v_2 (p_{T})$ of charged hadrons at forward rapidity 
appear to be systematically lower than previously published PHENIX 
results at midrapidity~\cite{PHENIX:2014yml} and consistent with 
expectations from the PHOBOS measurement~\cite{PHOBOS:2004vcu} of the 
rapidity dependence of charged hadrons.  A significant elliptic flow 
$v_2 (p_{T})$ was seen in the open-heavy-flavor-muon sample. A clear 
difference was observed between the flow of charged hadrons and the 
muons from heavy-flavor-hadron decays. This difference suggests that 
there is a kinematically-driven quark-mass dependence of the particle 
interactions with the QGP. The open-heavy-flavor muon $v_2$ is 
comparable in magnitude to previous PHENIX measurements of $v_2$ of 
open-heavy-flavor electrons at midrapidity~\cite{PHENIX:2006iih}. Taken 
together, the charged-hadron and the heavy-flavor-muon measurements 
suggest that there is no strong rapidity dependence in the collective 
flow of light and heavy-flavor hadrons in the measured rapidity range 
at RHIC energies.




\begin{acknowledgments}

We thank the staff of the Collider-Accelerator and Physics
Departments at Brookhaven National Laboratory and the staff of
the other PHENIX participating institutions for their vital
contributions.  
We acknowledge support from the Office of Nuclear Physics in the
Office of Science of the Department of Energy,
the National Science Foundation,
Abilene Christian University Research Council,
Research Foundation of SUNY, and
Dean of the College of Arts and Sciences, Vanderbilt University
(USA),
Ministry of Education, Culture, Sports, Science, and Technology
and the Japan Society for the Promotion of Science (Japan),
Natural Science Foundation of China (People's Republic of China),
Croatian Science Foundation and
Ministry of Science and Education (Croatia),
Ministry of Education, Youth and Sports (Czech Republic),
Centre National de la Recherche Scientifique, Commissariat
{\`a} l'{\'E}nergie Atomique, and Institut National de Physique
Nucl{\'e}aire et de Physique des Particules (France),
J. Bolyai Research Scholarship, EFOP, HUN-REN ATOMKI, NKFIH,
and OTKA (Hungary),
Department of Atomic Energy and Department of Science and Technology (India),
Israel Science Foundation (Israel),
Basic Science Research and SRC(CENuM) Programs through NRF
funded by the Ministry of Education and the Ministry of
Science and ICT (Korea),
Ministry of Education and Science, Russian Academy of Sciences,
Federal Agency of Atomic Energy (Russia),
VR and Wallenberg Foundation (Sweden),
University of Zambia, the Government of the Republic of Zambia (Zambia),
the U.S. Civilian Research and Development Foundation for the
Independent States of the Former Soviet Union,
the Hungarian American Enterprise Scholarship Fund,
the US-Hungarian Fulbright Foundation,
and the US-Israel Binational Science Foundation.

\end{acknowledgments}

\section*{DATA AVAILABILITY}

The data that support the findings of this article are not publicly available.
The values in the plots and tables associated with this article are stored in
HEPData~\cite{hepdata}.


\begin{thebibliography}{31}%
\makeatletter
\providecommand \@ifxundefined [1]{%
 \@ifx{#1\undefined}
}%
\providecommand \@ifnum [1]{%
 \ifnum #1\expandafter \@firstoftwo
 \else \expandafter \@secondoftwo
 \fi
}%
\providecommand \@ifx [1]{%
 \ifx #1\expandafter \@firstoftwo
 \else \expandafter \@secondoftwo
 \fi
}%
\providecommand \natexlab [1]{#1}%
\providecommand \enquote  [1]{``#1''}%
\providecommand \bibnamefont  [1]{#1}%
\providecommand \bibfnamefont [1]{#1}%
\providecommand \citenamefont [1]{#1}%
\providecommand \href@noop [0]{\@secondoftwo}%
\providecommand \href [0]{\begingroup \@sanitize@url \@href}%
\providecommand \@href[1]{\@@startlink{#1}\@@href}%
\providecommand \@@href[1]{\endgroup#1\@@endlink}%
\providecommand \@sanitize@url [0]{\catcode `\\12\catcode `\$12\catcode
  `\&12\catcode `\#12\catcode `\^12\catcode `\_12\catcode `\%12\relax}%
\providecommand \@@startlink[1]{}%
\providecommand \@@endlink[0]{}%
\providecommand \url  [0]{\begingroup\@sanitize@url \@url }%
\providecommand \@url [1]{\endgroup\@href {#1}{\urlprefix }}%
\providecommand \urlprefix  [0]{URL }%
\providecommand \Eprint [0]{\href }%
\providecommand \doibase [0]{https://doi.org/}%
\providecommand \selectlanguage [0]{\@gobble}%
\providecommand \bibinfo  [0]{\@secondoftwo}%
\providecommand \bibfield  [0]{\@secondoftwo}%
\providecommand \translation [1]{[#1]}%
\providecommand \BibitemOpen [0]{}%
\providecommand \bibitemStop [0]{}%
\providecommand \bibitemNoStop [0]{.\EOS\space}%
\providecommand \EOS [0]{\spacefactor3000\relax}%
\providecommand \BibitemShut  [1]{\csname bibitem#1\endcsname}%
\let\auto@bib@innerbib\@empty
\bibitem [{\citenamefont {Adcox}\ \emph {et~al.}(2005)\citenamefont {Adcox}
  \emph {et~al.}}]{PHENIX:2004vcz}%
  \BibitemOpen
  \bibfield  {author} {\bibinfo {author} {\bibfnamefont {K.}~\bibnamefont
  {Adcox}} \emph {et~al.} (\bibinfo {collaboration} {PHENIX Collaboration}),\
  }\bibfield  {title} {\bibinfo {title} {{Formation of dense partonic matter in
  relativistic nucleus-nucleus collisions at RHIC: Experimental evaluation by
  the PHENIX collaboration}},\ }\href
  {https://doi.org/10.1016/j.nuclphysa.2005.03.086} {\bibfield  {journal}
  {\bibinfo  {journal} {Nucl. Phys. A}\ }\textbf {\bibinfo {volume} {757}},\
  \bibinfo {pages} {184} (\bibinfo {year} {2005})}\BibitemShut {NoStop}%
\bibitem [{\citenamefont {Arsene}\ \emph {et~al.}(2005)\citenamefont {Arsene}
  \emph {et~al.}}]{BRAHMS:2004adc}%
  \BibitemOpen
  \bibfield  {author} {\bibinfo {author} {\bibfnamefont {I.}~\bibnamefont
  {Arsene}} \emph {et~al.} (\bibinfo {collaboration} {BRAHMS Collaboration}),\
  }\bibfield  {title} {\bibinfo {title} {{Quark gluon plasma and color glass
  condensate at RHIC? The Perspective from the BRAHMS experiment}},\ }\href
  {https://doi.org/10.1016/j.nuclphysa.2005.02.130} {\bibfield  {journal}
  {\bibinfo  {journal} {Nucl. Phys. A}\ }\textbf {\bibinfo {volume} {757}},\
  \bibinfo {pages} {1} (\bibinfo {year} {2005})}\BibitemShut {NoStop}%
\bibitem [{\citenamefont {Back}\ \emph
  {et~al.}(2005{\natexlab{a}})\citenamefont {Back} \emph
  {et~al.}}]{PHOBOS:2004zne}%
  \BibitemOpen
  \bibfield  {author} {\bibinfo {author} {\bibfnamefont {B.~B.}\ \bibnamefont
  {Back}} \emph {et~al.} (\bibinfo {collaboration} {PHOBOS Collaboration}),\
  }\bibfield  {title} {\bibinfo {title} {{The PHOBOS perspective on discoveries
  at RHIC}},\ }\href {https://doi.org/10.1016/j.nuclphysa.2005.03.084}
  {\bibfield  {journal} {\bibinfo  {journal} {Nucl. Phys. A}\ }\textbf
  {\bibinfo {volume} {757}},\ \bibinfo {pages} {28} (\bibinfo {year}
  {2005}{\natexlab{a}})}\BibitemShut {NoStop}%
\bibitem [{\citenamefont {Adams}\ \emph {et~al.}(2005)\citenamefont {Adams}
  \emph {et~al.}}]{STAR:2005gfr}%
  \BibitemOpen
  \bibfield  {author} {\bibinfo {author} {\bibfnamefont {J.}~\bibnamefont
  {Adams}} \emph {et~al.} (\bibinfo {collaboration} {STAR Collaboration}),\
  }\bibfield  {title} {\bibinfo {title} {{Experimental and theoretical
  challenges in the search for the quark gluon plasma: The STAR Collaboration's
  critical assessment of the evidence from RHIC collisions}},\ }\href
  {https://doi.org/10.1016/j.nuclphysa.2005.03.085} {\bibfield  {journal}
  {\bibinfo  {journal} {Nucl. Phys. A}\ }\textbf {\bibinfo {volume} {757}},\
  \bibinfo {pages} {102} (\bibinfo {year} {2005})}\BibitemShut {NoStop}%
\bibitem [{\citenamefont {Shuryak}(2005)}]{Shuryak:2004cy}%
  \BibitemOpen
  \bibfield  {author} {\bibinfo {author} {\bibfnamefont {E.~V.}\ \bibnamefont
  {Shuryak}},\ }\bibfield  {title} {\bibinfo {title} {{What RHIC experiments
  and theory tell us about properties of quark-gluon plasma?}},\ }\href
  {https://doi.org/10.1016/j.nuclphysa.2004.10.022} {\bibfield  {journal}
  {\bibinfo  {journal} {Nucl. Phys. A}\ }\textbf {\bibinfo {volume} {750}},\
  \bibinfo {pages} {64} (\bibinfo {year} {2005})}\BibitemShut {NoStop}%
\bibitem [{\citenamefont {Cacciari}\ \emph {et~al.}(2005)\citenamefont
  {Cacciari}, \citenamefont {Nason},\ and\ \citenamefont
  {Vogt}}]{Cacciari:2005rk}%
  \BibitemOpen
  \bibfield  {author} {\bibinfo {author} {\bibfnamefont {M.}~\bibnamefont
  {Cacciari}}, \bibinfo {author} {\bibfnamefont {P.}~\bibnamefont {Nason}},\
  and\ \bibinfo {author} {\bibfnamefont {R.}~\bibnamefont {Vogt}},\ }\bibfield
  {title} {\bibinfo {title} {{QCD predictions for charm and bottom production
  at RHIC}},\ }\href {https://doi.org/10.1103/PhysRevLett.95.122001} {\bibfield
   {journal} {\bibinfo  {journal} {Phys. Rev. Lett.}\ }\textbf {\bibinfo
  {volume} {95}},\ \bibinfo {pages} {122001} (\bibinfo {year}
  {2005})}\BibitemShut {NoStop}%
\bibitem [{\citenamefont {Dong}\ \emph {et~al.}(2019)\citenamefont {Dong},
  \citenamefont {Lee},\ and\ \citenamefont
  {Rapp}}]{annurev:annurev-nucl-101918-023806}%
  \BibitemOpen
  \bibfield  {author} {\bibinfo {author} {\bibfnamefont {X.}~\bibnamefont
  {Dong}}, \bibinfo {author} {\bibfnamefont {Y.-J.}\ \bibnamefont {Lee}},\ and\
  \bibinfo {author} {\bibfnamefont {R.}~\bibnamefont {Rapp}},\ }\bibfield
  {title} {\bibinfo {title} {Open heavy-flavor production in heavy-ion
  collisions},\ }\href
  {https://doi.org/https://doi.org/10.1146/annurev-nucl-101918-023806}
  {\bibfield  {journal} {\bibinfo  {journal} {Annual Review of Nuclear and
  Particle Science}\ }\textbf {\bibinfo {volume} {69}},\ \bibinfo {pages} {417}
  (\bibinfo {year} {2019})}\BibitemShut {NoStop}%
\bibitem [{\citenamefont {Adare}\ \emph {et~al.}(2016)\citenamefont {Adare}
  \emph {et~al.}}]{PHENIX:2015ynp}%
  \BibitemOpen
  \bibfield  {author} {\bibinfo {author} {\bibfnamefont {A.}~\bibnamefont
  {Adare}} \emph {et~al.} (\bibinfo {collaboration} {PHENIX Collaboration}),\
  }\bibfield  {title} {\bibinfo {title} {{Single electron yields from
  semileptonic charm and bottom hadron decays in Au$+$Au collisions at
  $\sqrt{s_{NN}}=200$ GeV}},\ }\href
  {https://doi.org/10.1103/PhysRevC.93.034904} {\bibfield  {journal} {\bibinfo
  {journal} {Phys. Rev. C}\ }\textbf {\bibinfo {volume} {93}},\ \bibinfo
  {pages} {034904} (\bibinfo {year} {2016})}\BibitemShut {NoStop}%
\bibitem [{\citenamefont {Abdulameer}\ \emph {et~al.}(2024)\citenamefont
  {Abdulameer} \emph {et~al.}}]{PHENIX:2022wim}%
  \BibitemOpen
  \bibfield  {author} {\bibinfo {author} {\bibfnamefont {N.~J.}\ \bibnamefont
  {Abdulameer}} \emph {et~al.} (\bibinfo {collaboration} {PHENIX
  Collaboration}),\ }\bibfield  {title} {\bibinfo {title} {{Charm- and
  bottom-quark production in Au+Au collisions at $\sqrt{s_{NN}}=200$ GeV}},\
  }\href {https://doi.org/10.1103/PhysRevC.109.044907} {\bibfield  {journal}
  {\bibinfo  {journal} {Phys. Rev. C}\ }\textbf {\bibinfo {volume} {109}},\
  \bibinfo {pages} {044907} (\bibinfo {year} {2024})}\BibitemShut {NoStop}%
\bibitem [{\citenamefont {Mustafa}\ \emph {et~al.}(1998)\citenamefont
  {Mustafa}, \citenamefont {Pal}, \citenamefont {Srivastava},\ and\
  \citenamefont {Thoma}}]{Mustafa:1997pm}%
  \BibitemOpen
  \bibfield  {author} {\bibinfo {author} {\bibfnamefont {M.~G.}\ \bibnamefont
  {Mustafa}}, \bibinfo {author} {\bibfnamefont {D.}~\bibnamefont {Pal}},
  \bibinfo {author} {\bibfnamefont {D.~K.}\ \bibnamefont {Srivastava}},\ and\
  \bibinfo {author} {\bibfnamefont {M.}~\bibnamefont {Thoma}},\ }\bibfield
  {title} {\bibinfo {title} {{Radiative energy loss of heavy quarks in a quark
  gluon plasma}},\ }\href {https://doi.org/10.1016/S0370-2693(98)00429-8}
  {\bibfield  {journal} {\bibinfo  {journal} {Phys. Lett. B}\ }\textbf
  {\bibinfo {volume} {428}},\ \bibinfo {pages} {234} (\bibinfo {year}
  {1998})}\BibitemShut {NoStop}%
\bibitem [{\citenamefont {Dokshitzer}\ and\ \citenamefont
  {Kharzeev}(2001)}]{Dokshitzer:2001zm}%
  \BibitemOpen
  \bibfield  {author} {\bibinfo {author} {\bibfnamefont {Y.~L.}\ \bibnamefont
  {Dokshitzer}}\ and\ \bibinfo {author} {\bibfnamefont {D.~E.}\ \bibnamefont
  {Kharzeev}},\ }\bibfield  {title} {\bibinfo {title} {{Heavy quark colorimetry
  of QCD matter}},\ }\href {https://doi.org/10.1016/S0370-2693(01)01130-3}
  {\bibfield  {journal} {\bibinfo  {journal} {Phys. Lett. B}\ }\textbf
  {\bibinfo {volume} {519}},\ \bibinfo {pages} {199} (\bibinfo {year}
  {2001})}\BibitemShut {NoStop}%
\bibitem [{\citenamefont {Meistrenko}\ \emph {et~al.}(2013)\citenamefont
  {Meistrenko}, \citenamefont {Peshier}, \citenamefont {Uphoff},\ and\
  \citenamefont {Greiner}}]{Meistrenko:2012ju}%
  \BibitemOpen
  \bibfield  {author} {\bibinfo {author} {\bibfnamefont {A.}~\bibnamefont
  {Meistrenko}}, \bibinfo {author} {\bibfnamefont {A.}~\bibnamefont {Peshier}},
  \bibinfo {author} {\bibfnamefont {J.}~\bibnamefont {Uphoff}},\ and\ \bibinfo
  {author} {\bibfnamefont {C.}~\bibnamefont {Greiner}},\ }\bibfield  {title}
  {\bibinfo {title} {{Collisional energy loss of heavy quarks}},\ }\href
  {https://doi.org/10.1016/j.nuclphysa.2013.02.012} {\bibfield  {journal}
  {\bibinfo  {journal} {Nucl. Phys. A}\ }\textbf {\bibinfo {volume} {901}},\
  \bibinfo {pages} {51} (\bibinfo {year} {2013})}\BibitemShut {NoStop}%
\bibitem [{\citenamefont {Cao}\ \emph {et~al.}(2019)\citenamefont {Cao} \emph
  {et~al.}}]{Cao:2018ews}%
  \BibitemOpen
  \bibfield  {author} {\bibinfo {author} {\bibfnamefont {S.}~\bibnamefont
  {Cao}} \emph {et~al.},\ }\bibfield  {title} {\bibinfo {title} {{Toward the
  determination of heavy-quark transport coefficients in quark-gluon plasma}},\
  }\href {https://doi.org/10.1103/PhysRevC.99.054907} {\bibfield  {journal}
  {\bibinfo  {journal} {Phys. Rev. C}\ }\textbf {\bibinfo {volume} {99}},\
  \bibinfo {pages} {054907} (\bibinfo {year} {2019})}\BibitemShut {NoStop}%
\bibitem [{\citenamefont {He}\ and\ \citenamefont {Rapp}(2020)}]{He:2019vgs}%
  \BibitemOpen
  \bibfield  {author} {\bibinfo {author} {\bibfnamefont {M.}~\bibnamefont
  {He}}\ and\ \bibinfo {author} {\bibfnamefont {R.}~\bibnamefont {Rapp}},\
  }\bibfield  {title} {\bibinfo {title} {{Hadronization and Charm-Hadron Ratios
  in Heavy-Ion Collisions}},\ }\href
  {https://doi.org/10.1103/PhysRevLett.124.042301} {\bibfield  {journal}
  {\bibinfo  {journal} {Phys. Rev. Lett.}\ }\textbf {\bibinfo {volume} {124}},\
  \bibinfo {pages} {042301} (\bibinfo {year} {2020})}\BibitemShut {NoStop}%
\bibitem [{\citenamefont {Adare}\ \emph {et~al.}(2007)\citenamefont {Adare}
  \emph {et~al.}}]{PHENIX:2006iih}%
  \BibitemOpen
  \bibfield  {author} {\bibinfo {author} {\bibfnamefont {A.}~\bibnamefont
  {Adare}} \emph {et~al.} (\bibinfo {collaboration} {PHENIX Collaboration}),\
  }\bibfield  {title} {\bibinfo {title} {{Energy Loss and Flow of Heavy Quarks
  in Au+Au Collisions at $\sqrt{s_{NN}}=200$ GeV}},\ }\href
  {https://doi.org/10.1103/PhysRevLett.98.172301} {\bibfield  {journal}
  {\bibinfo  {journal} {Phys. Rev. Lett.}\ }\textbf {\bibinfo {volume} {98}},\
  \bibinfo {pages} {172301} (\bibinfo {year} {2007})}\BibitemShut {NoStop}%
\bibitem [{\citenamefont {Adamczyk}\ \emph {et~al.}(2017)\citenamefont
  {Adamczyk} \emph {et~al.}}]{STAR:2017kkh}%
  \BibitemOpen
  \bibfield  {author} {\bibinfo {author} {\bibfnamefont {L.}~\bibnamefont
  {Adamczyk}} \emph {et~al.} (\bibinfo {collaboration} {STAR Collaboration}),\
  }\bibfield  {title} {\bibinfo {title} {{Measurement of $D^0$ Azimuthal
  Anisotropy at Midrapidity in Au+Au Collisions at $\sqrt{s_{NN}}$=200 GeV}},\
  }\href {https://doi.org/10.1103/PhysRevLett.118.212301} {\bibfield  {journal}
  {\bibinfo  {journal} {Nucl. Instrum. Methods Phys. Res., Sec. A}\ }\textbf
  {\bibinfo {volume} {118}},\ \bibinfo {pages} {212301} (\bibinfo {year}
  {2017})}\BibitemShut {NoStop}%
\bibitem [{\citenamefont {Sirunyan}\ \emph {et~al.}(2021)\citenamefont
  {Sirunyan} \emph {et~al.}}]{CMS:2020bnz}%
  \BibitemOpen
  \bibfield  {author} {\bibinfo {author} {\bibfnamefont {A.~M.}\ \bibnamefont
  {Sirunyan}} \emph {et~al.} (\bibinfo {collaboration} {CMS Collaboration}),\
  }\bibfield  {title} {\bibinfo {title} {{Measurement of prompt
  ${\mathrm{D^0}}$ and ${\mathrm{\overline{D}}{}^0}$ meson azimuthal anisotropy
  and search for strong electric fields in PbPb collisions at
  $\sqrt{s_\mathrm{NN}} =$ 5.02 TeV}},\ }\href
  {https://doi.org/10.1016/j.physletb.2021.136253} {\bibfield  {journal}
  {\bibinfo  {journal} {Phys. Lett. B}\ }\textbf {\bibinfo {volume} {816}},\
  \bibinfo {pages} {136253} (\bibinfo {year} {2021})}\BibitemShut {NoStop}%
\bibitem [{\citenamefont {Morrison}\ \emph {et~al.}(1998)\citenamefont
  {Morrison} \emph {et~al.}}]{PHENIX:1998vmi}%
  \BibitemOpen
  \bibfield  {author} {\bibinfo {author} {\bibfnamefont {D.~P.}\ \bibnamefont
  {Morrison}} \emph {et~al.} (\bibinfo {collaboration} {PHENIX
  Collaboration}),\ }\bibfield  {title} {\bibinfo {title} {{The PHENIX
  experiment at RHIC}},\ }\href {https://doi.org/10.1016/S0375-9474(98)00390-X}
  {\bibfield  {journal} {\bibinfo  {journal} {Nucl. Phys. A}\ }\textbf
  {\bibinfo {volume} {638}},\ \bibinfo {pages} {565} (\bibinfo {year}
  {1998})}\BibitemShut {NoStop}%
\bibitem [{\citenamefont {Allen}\ \emph {et~al.}(2003)\citenamefont {Allen}
  \emph {et~al.}}]{ALLEN2003549}%
  \BibitemOpen
  \bibfield  {author} {\bibinfo {author} {\bibfnamefont {M.}~\bibnamefont
  {Allen}} \emph {et~al.} (\bibinfo {collaboration} {PHENIX Collaboration}),\
  }\bibfield  {title} {\bibinfo {title} {Phenix inner detectors},\ }\href
  {https://doi.org/https://doi.org/10.1016/S0168-9002(02)01956-3} {\bibfield
  {journal} {\bibinfo  {journal} {Nucl. Instrum. Methods Phys. Res., Sec. A}\
  }\textbf {\bibinfo {volume} {499}},\ \bibinfo {pages} {549} (\bibinfo {year}
  {2003})}\BibitemShut {NoStop}%
\bibitem [{\citenamefont {Nouicer}(2007)}]{Nouicer:2007rb}%
  \BibitemOpen
  \bibfield  {author} {\bibinfo {author} {\bibfnamefont {R.}~\bibnamefont
  {Nouicer}} (\bibinfo {collaboration} {PHENIX Collaboration}),\ }\bibfield
  {title} {\bibinfo {title} {{PHENIX Upgrade: Novel Stripixel Detector for
  Heavy Quark Detection and Proton Spin Structure Measurements at RHIC
  Energies}},\ }\href {https://doi.org/10.1016/j.nimb.2007.04.265} {\bibfield
  {journal} {\bibinfo  {journal} {Nucl. Instrum. Methods Phys. Res., Sec. B}\
  }\textbf {\bibinfo {volume} {261}},\ \bibinfo {pages} {1067} (\bibinfo {year}
  {2007})}\BibitemShut {NoStop}%
\bibitem [{\citenamefont {Akikawa}\ \emph {et~al.}(2003)\citenamefont {Akikawa}
  \emph {et~al.}}]{Akikawa:2003zs}%
  \BibitemOpen
  \bibfield  {author} {\bibinfo {author} {\bibfnamefont {H.}~\bibnamefont
  {Akikawa}} \emph {et~al.} (\bibinfo {collaboration} {PHENIX Collaboration}),\
  }\bibfield  {title} {\bibinfo {title} {{PHENIX muon arms}},\ }\href
  {https://doi.org/10.1016/S0168-9002(02)01955-1} {\bibfield  {journal}
  {\bibinfo  {journal} {Nucl. Instrum. Methods Phys. Res., Sec. A}\ }\textbf
  {\bibinfo {volume} {499}},\ \bibinfo {pages} {537} (\bibinfo {year}
  {2003})}\BibitemShut {NoStop}%
\bibitem [{\citenamefont {Aidala}\ \emph {et~al.}(2014)\citenamefont {Aidala}
  \emph {et~al.}}]{AIDALA201444}%
  \BibitemOpen
  \bibfield  {author} {\bibinfo {author} {\bibfnamefont {C.}~\bibnamefont
  {Aidala}} \emph {et~al.} (\bibinfo {collaboration} {PHENIX Collaboration}),\
  }\bibfield  {title} {\bibinfo {title} {The phenix forward silicon vertex
  detector},\ }\href
  {https://doi.org/https://doi.org/10.1016/j.nima.2014.04.017} {\bibfield
  {journal} {\bibinfo  {journal} {Nucl. Instrum. Methods Phys. Res., Sec. A}\
  }\textbf {\bibinfo {volume} {755}},\ \bibinfo {pages} {44} (\bibinfo {year}
  {2014})}\BibitemShut {NoStop}%
\bibitem [{\citenamefont {Poskanzer}\ and\ \citenamefont
  {Voloshin}(1998)}]{Poskanzer:1998yz}%
  \BibitemOpen
  \bibfield  {author} {\bibinfo {author} {\bibfnamefont {A.~M.}\ \bibnamefont
  {Poskanzer}}\ and\ \bibinfo {author} {\bibfnamefont {S.~A.}\ \bibnamefont
  {Voloshin}},\ }\bibfield  {title} {\bibinfo {title} {{Methods for analyzing
  anisotropic flow in relativistic nuclear collisions}},\ }\href
  {https://doi.org/10.1103/PhysRevC.58.1671} {\bibfield  {journal} {\bibinfo
  {journal} {Phys. Rev. C}\ }\textbf {\bibinfo {volume} {58}},\ \bibinfo
  {pages} {1671} (\bibinfo {year} {1998})}\BibitemShut {NoStop}%
\bibitem [{\citenamefont {Bierlich}\ \emph {et~al.}(2022)\citenamefont
  {Bierlich} \emph {et~al.}}]{Bierlich:2022pfr}%
  \BibitemOpen
  \bibfield  {author} {\bibinfo {author} {\bibfnamefont {C.}~\bibnamefont
  {Bierlich}} \emph {et~al.},\ }\bibfield  {title} {\bibinfo {title} {{A
  comprehensive guide to the physics and usage of PYTHIA 8.3}},\ }\href
  {https://doi.org/10.21468/SciPostPhysCodeb.8} {\bibfield  {journal} {\bibinfo
   {journal} {SciPost Phys. Codebases}\ }\textbf {\bibinfo {volume} {2022}},\
  \bibinfo {pages} {8} (\bibinfo {year} {2022})}\BibitemShut {NoStop}%
\bibitem [{\citenamefont {Agostinelli}\ \emph {et~al.}(2003)\citenamefont
  {Agostinelli} \emph {et~al.}}]{GEANT4:2002zbu}%
  \BibitemOpen
  \bibfield  {author} {\bibinfo {author} {\bibfnamefont {S.}~\bibnamefont
  {Agostinelli}} \emph {et~al.} (\bibinfo {collaboration} {GEANT4
  Collaboration}),\ }\bibfield  {title} {\bibinfo {title} {{GEANT4-a simulation
  toolkit}},\ }\href {https://doi.org/10.1016/S0168-9002(03)01368-8} {\bibfield
   {journal} {\bibinfo  {journal} {Nucl. Instrum. Methods Phys. Res., Sec. A}\
  }\textbf {\bibinfo {volume} {506}},\ \bibinfo {pages} {250} (\bibinfo {year}
  {2003})}\BibitemShut {NoStop}%
\bibitem [{\citenamefont {Adare}\ \emph {et~al.}(2013)\citenamefont {Adare}
  \emph {et~al.}}]{PHENIX:2013kod}%
  \BibitemOpen
  \bibfield  {author} {\bibinfo {author} {\bibfnamefont {A.}~\bibnamefont
  {Adare}} \emph {et~al.} (\bibinfo {collaboration} {PHENIX Collaboration}),\
  }\bibfield  {title} {\bibinfo {title} {{Spectra and ratios of identified
  particles in Au+Au and $d$+Au collisions at $\sqrt{s_{NN}}=200$ GeV}},\
  }\href {https://doi.org/10.1103/PhysRevC.88.024906} {\bibfield  {journal}
  {\bibinfo  {journal} {Phys. Rev. C}\ }\textbf {\bibinfo {volume} {88}},\
  \bibinfo {pages} {024906} (\bibinfo {year} {2013})}\BibitemShut {NoStop}%
\bibitem [{\citenamefont {Arsene}\ \emph {et~al.}(2016)\citenamefont {Arsene}
  \emph {et~al.}}]{BRAHMS:2016klg}%
  \BibitemOpen
  \bibfield  {author} {\bibinfo {author} {\bibfnamefont {I.~C.}\ \bibnamefont
  {Arsene}} \emph {et~al.} (\bibinfo {collaboration} {BRAHMS Collaboration}),\
  }\bibfield  {title} {\bibinfo {title} {{Rapidity and centrality dependence of
  particle production for identified hadrons in Cu+Cu collisions at
  $\sqrt{s_{NN}} = 200$ GeV}},\ }\href
  {https://doi.org/10.1103/PhysRevC.94.014907} {\bibfield  {journal} {\bibinfo
  {journal} {Phys. Rev. C}\ }\textbf {\bibinfo {volume} {94}},\ \bibinfo
  {pages} {014907} (\bibinfo {year} {2016})}\BibitemShut {NoStop}%
\bibitem [{\citenamefont {Adare}\ \emph {et~al.}(2015)\citenamefont {Adare}
  \emph {et~al.}}]{PHENIX:2014yml}%
  \BibitemOpen
  \bibfield  {author} {\bibinfo {author} {\bibfnamefont {A.}~\bibnamefont
  {Adare}} \emph {et~al.} (\bibinfo {collaboration} {PHENIX Collaboration}),\
  }\bibfield  {title} {\bibinfo {title} {{Systematic Study of Azimuthal
  Anisotropy in Cu$+$Cu and Au$+$Au Collisions at $\sqrt{s_{_{NN}}} = 62.4$ and
  200 GeV}},\ }\href {https://doi.org/10.1103/PhysRevC.92.034913} {\bibfield
  {journal} {\bibinfo  {journal} {Phys. Rev. C}\ }\textbf {\bibinfo {volume}
  {92}},\ \bibinfo {pages} {034913} (\bibinfo {year} {2015})}\BibitemShut
  {NoStop}%
\bibitem [{\citenamefont {Back}\ \emph
  {et~al.}(2005{\natexlab{b}})\citenamefont {Back} \emph
  {et~al.}}]{PHOBOS:2004vcu}%
  \BibitemOpen
  \bibfield  {author} {\bibinfo {author} {\bibfnamefont {B.~B.}\ \bibnamefont
  {Back}} \emph {et~al.} (\bibinfo {collaboration} {PHOBOS Collaboration}),\
  }\bibfield  {title} {\bibinfo {title} {{Centrality and pseudorapidity
  dependence of elliptic flow for charged hadrons in Au+Au collisions at
  $\sqrt{s_{NN}}=200$ GeV}},\ }\href
  {https://doi.org/10.1103/PhysRevC.72.051901} {\bibfield  {journal} {\bibinfo
  {journal} {Phys. Rev. C}\ }\textbf {\bibinfo {volume} {72}},\ \bibinfo
  {pages} {051901} (\bibinfo {year} {2005}{\natexlab{b}})}\BibitemShut
  {NoStop}%
\bibitem [{\citenamefont {Adler}\ \emph {et~al.}(2004)\citenamefont {Adler}
  \emph {et~al.}}]{PHENIX:2003iij}%
  \BibitemOpen
  \bibfield  {author} {\bibinfo {author} {\bibfnamefont {S.~S.}\ \bibnamefont
  {Adler}} \emph {et~al.} (\bibinfo {collaboration} {PHENIX Collaboration}),\
  }\bibfield  {title} {\bibinfo {title} {{Identified charged particle spectra
  and yields in Au+Au collisions at $\sqrt{s}=200$ GeV}},\ }\href
  {https://doi.org/10.1103/PhysRevC.69.034909} {\bibfield  {journal} {\bibinfo
  {journal} {Phys. Rev. C}\ }\textbf {\bibinfo {volume} {69}},\ \bibinfo
  {pages} {034909} (\bibinfo {year} {2004})}\BibitemShut {NoStop}%
\bibitem [{hep()}]{hepdata}%
  \BibitemOpen
  \href@noop {} {}\bibinfo {howpublished}
  {https://www.hepdata.net/record/159542}\BibitemShut {NoStop}%
\end{thebibliography}

%
 
\end{document}